\documentclass[pre,twocolumn,groupedaddress,showpacs,showkeys,amsmath]{revtex4-1}

\usepackage[pdftex]{graphicx}
\usepackage{dcolumn}
\usepackage{bm}
\usepackage{amsfonts}
\usepackage{subfigure}
\usepackage{epsfig}
\graphicspath{{./PDF/}{./EPS/}}

\begin{document}


\title{Convex Hulls of Random Walks: Large-Deviation Properties}

\author{Gunnar Claussen}
\email{gunnar.claussen@uni-oldenburg.de}
\author{Alexander K. Hartmann}
\email{alexander.hartmann@uni-oldenburg.de}
\affiliation{
 Institut f\"ur Physik, Universit\"at Oldenburg, 
Carl-von-Ossietzky-Stra{\ss}e 9--11, 26111 Oldenburg, Germany}
\author{Satya N. Majumdar}
\email{satya.majumdar@u-psud.fr}
\affiliation{Laboratoire de Physique Th\'eorique et Mod\`eles Statistiques (UMR 
8626 du CNRS), 
Universit\'e de Paris-Sud, B\^atiment 100, 91405 Orsay Cedex, France.
}

\date{\today}


\begin{abstract}
We study the convex hull of the set of points visited by a two-dimensional
random walker of $T$ discrete time steps. Two natural observables that
characterize the convex hull in two dimensions are its perimeter $L$ and
area $A$.
While the mean perimeter $\langle L \rangle$ and the mean area 
$\langle A \rangle$ have been studied before, analytically and numerically, and exact results are
known for large $T$ (Brownian motion limit),
little is known about the full distributions $P(A)$ and $P(L)$. 
In this paper, we provide numerical results for these distributions.
We use a sophisticated large-deviation
approach that allows us to study the distributions over a larger range of
the support, where the probabilities $P(A)$ and $P(L)$ are as small
as $10^{-300}$. We analyse (open) random walks as well as (closed) 
Brownian bridges 
on the two-dimensional discrete grid as well as in the two-dimensional plane.
The resulting distributions exhibit, for large $T$, a universal scaling behavior (independent
of the details of the jump distributions)   
as a function of $A/T$ and $L/\sqrt{T}$, respecively. We are also
able to obtain the rate function, describing rare events at the tails
of these distributions, via a numerical extrapolation scheme
and find a linear and square dependence as a function of
the rescaled perimeter and the rescaled area, respectively. 
\end{abstract}

\pacs{02.50.-r,75.40.Mg,89.75.Da}
\keywords{random walk, convex hull, large-deviation properties, rare events}
\maketitle


\section{Introduction \label{sec:introduction}}
Random walks, originally introduced in 1921 by G. 
P\'olya \cite{Polya1921}, have since been a vital field of research. 
They are ubiquitous models for physical, biological and social
processes \cite{vanKampen1992,bergHC1993,hughes1996}. 
Example applications from biology include self-propelled motion of 
bacteria and the diffusion of nutrients \cite{bergHC1993}, as well as 
animal motion in general \cite{Bovet1988,Bartumeus2005} or during 
the marking of territories or description of home ranges 
\cite{mohr1947,worton1995,Giuggioli2011}. 
For the latter case a strong increase of the amount of experimentally available 
data ocurred after the introduction of
automated radio/GPS tagging of animals \cite{kenward1987,landguth2010}.
The usage of minimum convex polygons, called \emph{convex hulls}, bordering the
trace of an animal \cite{mohr1947,worton1987} 
 is a simple yet versatile \cite{boyle2009} way to describe
the home range and can be used for any type of (random-walk) data. 
 In two dimensions, the convex hull of a point 
set is the minimum subset whose elements form a convex polygon in 
such a way that (a) all points of the set and (b) the connecting 
lines between all possible pairs lie inside the polygon. The convex hull 
is a suitable measure for time-discretized random walks because it 
can easily be attributed with geometrical quantities, i.e., hull area 
$A$ and perimeter $L$. Also, for numerical calculations, simple and 
already quite fast algorithms with running times 
$\mathcal{O}(N \cdot \log N)$ exist 
\cite{Graham1972, Bykat1978, Andrew1979}.

On the analytical side, much progress have been made for asymptotic 
results at long times, when 
a random walk (with a finite variance for step sizes) converges to 
the continuous-time Brownian motion (for a recent review see
Ref. \cite{Majumdar2010}). For a single 
Brownian motion of length $T$ in two dimensions, the mean 
perimeter~\cite{Takacs1980,Letac1993} and
the mean area~\cite{ElBachir1983} were computed long back. 

Recently, adapting Cauchy's formula~\cite{Cauchy1832} for convex curves in 
two dimensions to random curves, it was 
shown~\cite{Randon-Furling2009,Majumdar2010} that the problem of computing 
the mean perimeter and the mean area of the convex hull of an arbitrary 
two dimensional stochastic process can be mapped to computing the extremal 
statistics of the one dimensional component of the process. This procedure 
was successfully applied recently to compute the mean perimeter and the 
mean area of several two dimensional stochastic processes such as $N$ 
independent Brownian motions in 
$2$-d~\cite{Randon-Furling2009,Majumdar2010}, random acceleration process 
in $2$-d~\cite{Reymbaut2011}, $2$-d branching Brownian motions with absorption 
with applications to edpidemic outbreak~\cite{Dumonteil2013} and $2$-d 
anomalous diffusion processes~\cite{Lukovic2013}. Very recently,
this method was also successfully used to compute the exact mean perimeter
of the convex hull of a planar Brownian motion confined to a half-space~\cite{Chupeau2014}.
Finally, using different methods, the mean perimeter and the mean area of the convex hull of a single 
Brownian motion, but in arbitrary dimensions, have been 
computed recently in the mathematics literature~\cite{Eldan2011,Kabluchko2014}.

The question naturally arises regarding the higher moments or even
the full distribution of the perimeter and the area of the
convex hulls of two dimensional random walks.
Computing analytically even the second moment (and hence the variance),  
for the perimeter and the area of the convex hull of 
a single two dimensional Brownian motion, turns out
to be highly difficult~\cite{Snyder1993,Goldman1996}.
Since so far no analytical results are available concerning the full 
distributions, performing numerical studies is a natural approach.
Within
a straightforward implementation (``simple sampling''), 
one generates many times independently a random walk of $T$ 
steps and constructss the hull polygon and, subsequently computes
its area $A$ 
and perimeter $L$. Then one records histograms of these
quantities. Nevertheless, this approach
 only gives insight into a small portion of the 
actual distributions $P(A)$ and $P(L)$, i.e., configurations with 
$P \propto K^{-1}$, the inverse of the number $K$ of samples. This 
demands for specific calculations with respect to the 
large-deviation properties of this model. So, we use a particular 
large-deviation approach~\cite{align2002,Hartmann2011} which randomly 
alters steps of the random 
walk, compares the effected change in terms of either of the quantities 
and accepts the alteration according a Metropolis criterion 
involving an artifical Monte Carlo ``temperature'' 
$\Theta$. This is performed for various values of $\Theta$, giving access
to different ranges of the distribution under scrutiny.
Afterwards, the resulting distributions $P_{\Theta}(A)$ or 
$P_{\Theta}(L)$ can be merged according to a 
given scheme. In the end, we are able to obtain distributions $P(A)$ and 
$P(L)$ which cover a range in principle 
only limited by the numerical precision of our computers.

The rest of the  paper is organized as follows: 
Sec.\ \ref{sec:model} will introduce 
the used random walk model, the associated convex hull and then 
will shortly 
allude 
to the algorithms and some heuristics. Following this, in Sec.\  
\ref{sec:large_deviation_scheme} we will introduce the large-deviation 
method. Our results are presented in Sec. \ref{sec:results}, and we 
will finally conclude in Sec.\ \ref{sec:conclusions}, giving an 
outlook to possible future work.


\section{Random Walks, Convex Hulls and Algorithms \label{sec:model}}
A time-discretized random walk consists of $T$ step vectors 
$\vec{\delta}_i$, and the position $\vec{x}(\tau)$ at timestep 
$\tau < T$ is the sum of all steps up to $\tau$, i.e.:
\begin{equation}
 \vec{x}(\tau) = \vec{x}_0 + \sum^{\tau}_{i=1}\vec{\delta}_i
\end{equation}
The walk configuration itself is then the set 
$\mathcal{W} = \{\vec{\delta}_1, \vec{\delta}_1, ..., \vec{\delta}_T\}$ 
of steps \cite{Feller1950}. The step $\vec{\delta}_i = 
(\delta_{x,i}, \delta_{y,i})$ itself denotes a displacement of the 
particle by $\delta_{x,i}$ in $x$-direction and $\delta_{y,i}$
in $y$-direction. We consider two 
different types of steps:

\begin{enumerate}
 \item A time-discrete approximation to a Brownian walk, i.e., both 
$\delta_{x,i}$ and $\delta_{y,i}$ are, for each $i$, drawn randomly 
from a Gaussian distribution with zero mean and variance one.

 \item A walk with discrete steps corresponding to motion on a 
square lattice of spacing $J$. This corresponds to the four possible 
steps $\vec{x} \in \{(0,J),(0,-J),(-J,0),(J,0)\}$, with each of 
these possible up/down/left/right steps having probability 
$\frac{1}{4}$.
\end{enumerate}

Note that in the limit of long walks $T \rightarrow \infty$ Gaussian 
walks and discrete lattice walks should statisically agree when setting 
$J = \sqrt{2}$.  We also consider closed random walks which return to 
the origin, i.e., $\vec{x}(T) = \vec{x}(0)$:

\begin{enumerate}
 \item For the lattice case, we generate only the first half of the 
walk randomly. The second half is filled up with inverse steps 
$\vec{\delta}_{i+\frac{T}{2}} = -\vec{\delta}_i$. Finally the order of 
the steps is rando\-mized via swapping randomly selected pairs of 
steps.
 \item For the Gaussian case, we consider Brownian bridges 
\cite{Mansuy2008}, i.e., from the walk $\vec{x}(\tau)$ we 
construct
\begin{equation}
 \vec{x}_b(\tau) = \vec{x}(\tau) - \frac{\tau}{T} \vec{x}(T)
\end{equation}
 which fulfill $\vec{x}_b(T) = \vec{x}(0)$ and have all 
necessary  statistical properties of random walks.
\end{enumerate}

The convex hull $\mathcal{C} = \text{conv}(\mathcal{\tilde P})$ of a 
two-dimensional point set $\mathcal{\tilde P} = \{\tilde P_i\}, 
\tilde P_i \in \mathbb{R}^2$ 
is described through a convex set over $\mathcal{\tilde P}$. The points $P$ 
within $\mathcal{C}$ are given by all possible combinations 
$P = \sum \alpha_i \tilde P_i$ with $\tilde P_i \in \mathcal{\tilde 
P}$ and 
$\sum_i \alpha_i = 1$ and $\alpha_i \in \mathbb{R}^+_0$ (definition 
given according to \cite{Preparata1985}). This means:
\begin{enumerate}
 \item All points $P_i \in \mathcal{P}$ lie within $\mathcal{C}$.
 \item All lines $\overline{P_i P_j}; P_i, P_j \in \mathcal{P}$ also 
lie within $\mathcal{C}$.
\end{enumerate}

The boundary 
of the convex set 
is  a 
polygon which connects a subset $\mathcal{P}\subset \mathcal{\tilde P}$
of $H$ points from the point set, i.e.,  
$\mathcal{P}=\{P_0,P_1,\ldots,P_{H-1}\}$, with $P_i=(x_i,y_i)$ 
($i=0,\ldots,H-1$). The hull is attributed with area 
$A$ and perimeter $L$ according to (identifying $i=H$ with $i=0$):

\begin{align}
 A(\mathcal{C}) = \frac{1}{2} \sum_{i=0}^{H-1} (y_i + y_{i+1})\cdot (x_i-x_{i+1}) \\
 L(\mathcal{C}) = \sum_{i=0}^{H-1} \sqrt{(x_i-x_{i+1})^2 + (y_i-y_{i+1})^2}
\end{align}

For our work, we determined the polygons bordering convex hulls
(for which one uses shortly the  term ``convex hull'') numerically.
The most common convex hull algorithms operate in 
$\mathcal{O}(N \log N)$. We used Andrew's variant \cite{Andrew1979} 
of the ``Graham Scan'' algorithm \cite{Graham1972}, which constructs 
the convex hull by first drawing a dividing line through the point 
set and then by sorting out those points which don't form monotone chain 
of clockwise/counter-clockwise turns on each side of the line. In 
usual cases, the application of convex hull algorithms can be 
accelerated by usage of pre-selection heuristics, such as the 
one introduced by Akl and Toussaint \cite{Akl1978}. This heuristic looks 
up extreme points of the set (i.e., those of maximum and minimum $x$- 
and $y$-coordinates) and discards all points which lie inside 
the quadrilateral formed by these points. We use a custom refinement 
of this heuristic, which is based on iterating the heuristic under 
rotation of the coordinate origin, which eliminates another fraction 
of inert points per each iteration.

\section{Large-Devation Scheme \label{sec:large_deviation_scheme}}

For simple-sampling results, walk configurations $\mathcal{W}$ 
are generated randomly, and the according convex hulls $\mathcal{C}$ 
are calculated through the algorithm, resulting in a multitude values of 
$A$ and $L$. As mentioned above, obtaining histograms of these
values only gives access to the high probability regime, where
the convex- hull-properties of typical random walks are measured.
However, in order to obtain values of these quantities 
with especially low probabilities, allowing us the measure the distributions
$P(A)$ and $P(L)$ over a large range of the supoort, 
a certain Markov-Chain Monte 
Carlo (MCMC) scheme can be used \cite{align2002,Hartmann2011}.

\begin{figure}[ht!p]
\subfigure[ ]{\includegraphics[width=0.49\linewidth]{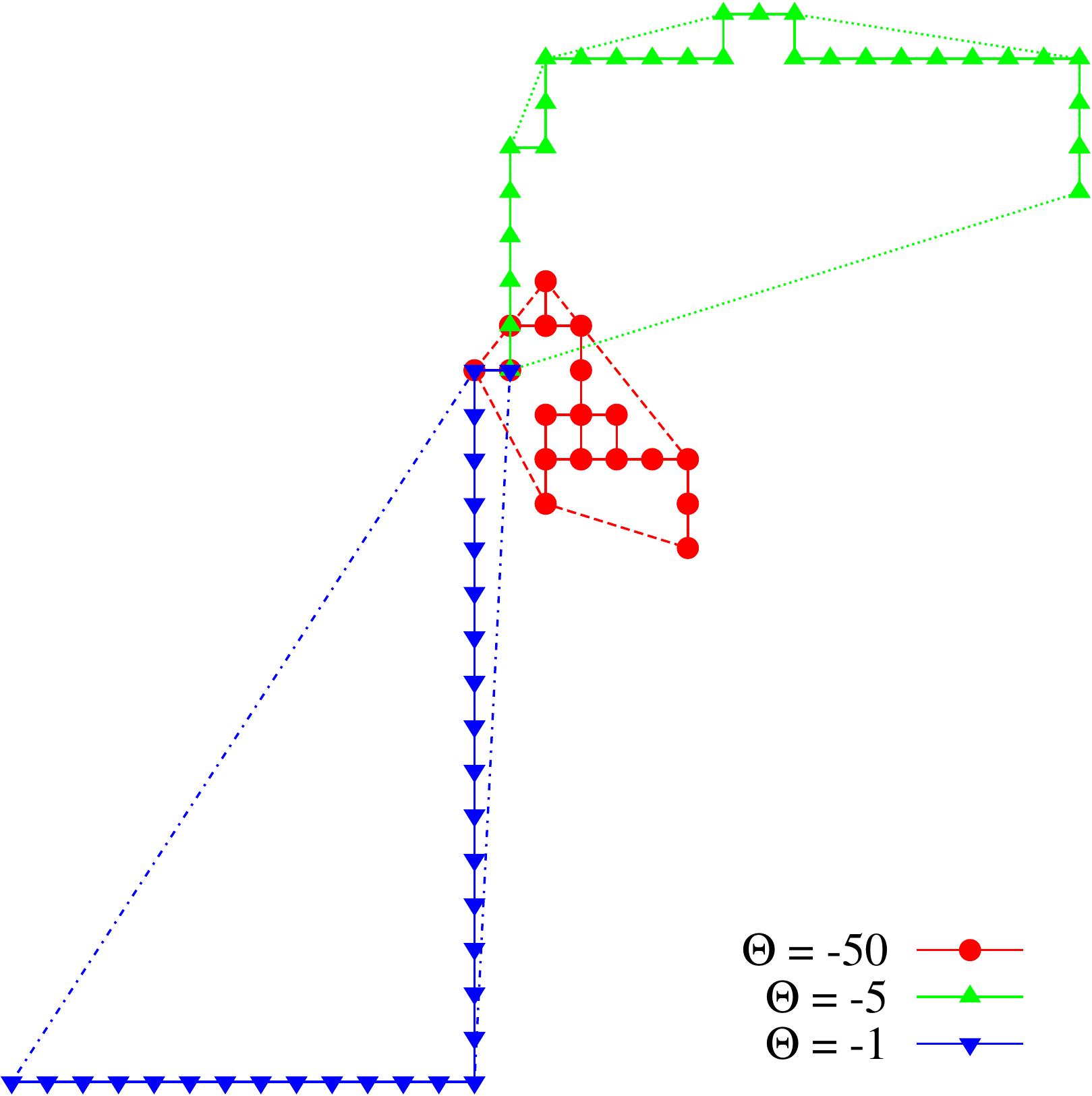}}
\subfigure[ ]{\includegraphics[width=0.49\linewidth]{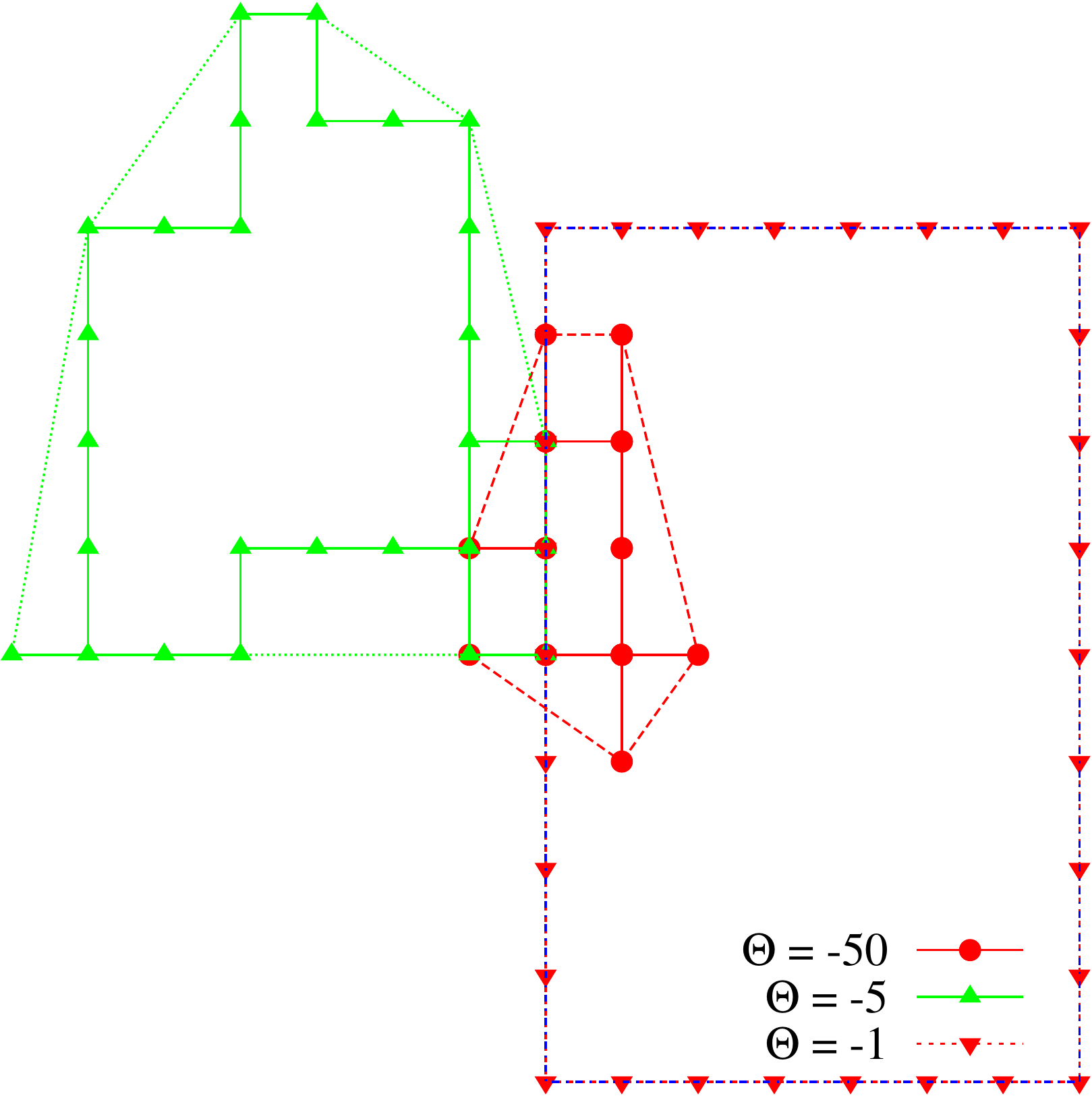}}
\subfigure[ ]{\includegraphics[width=0.49\linewidth]{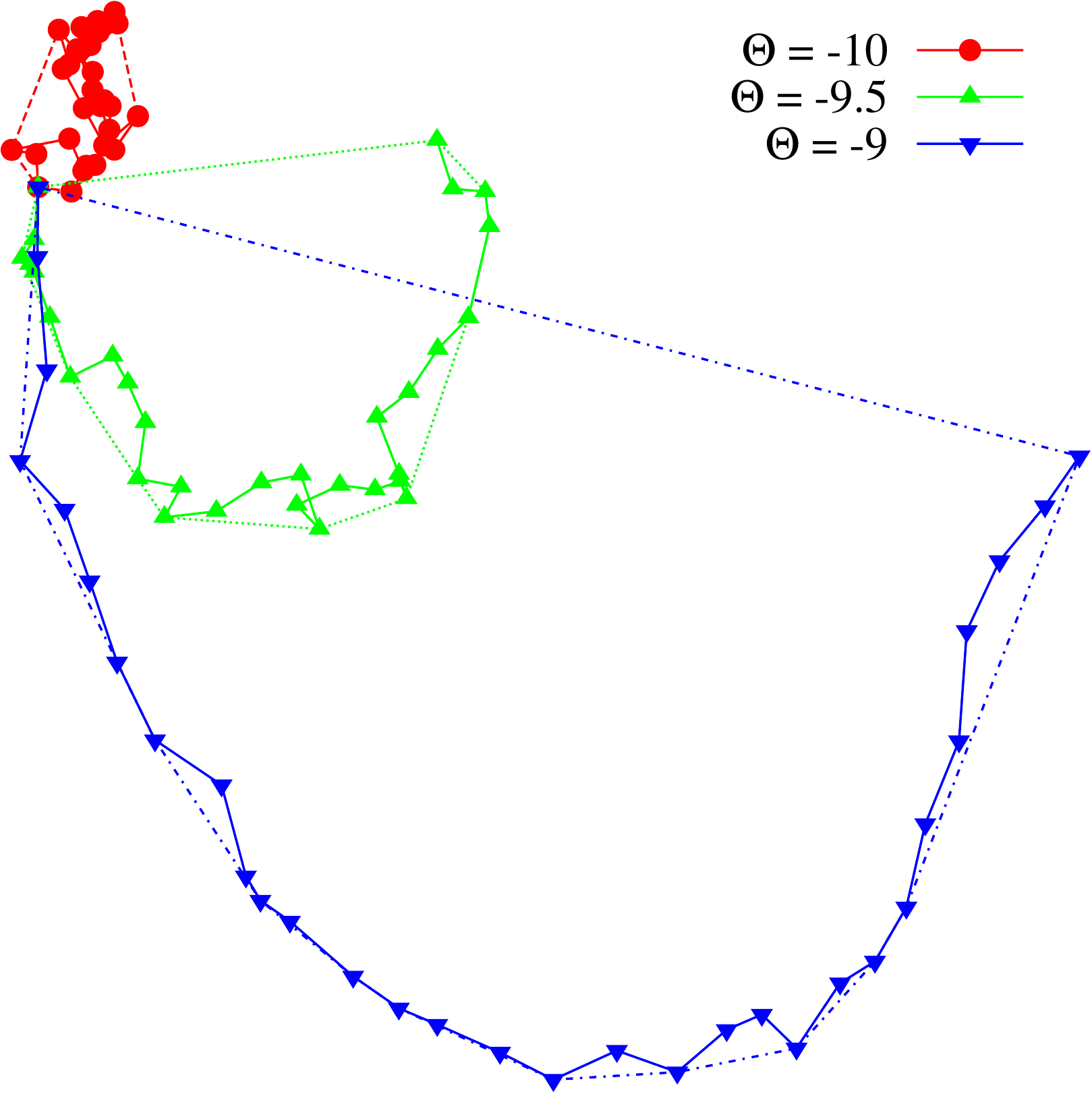}}
\subfigure[ ]{\includegraphics[width=0.49\linewidth]{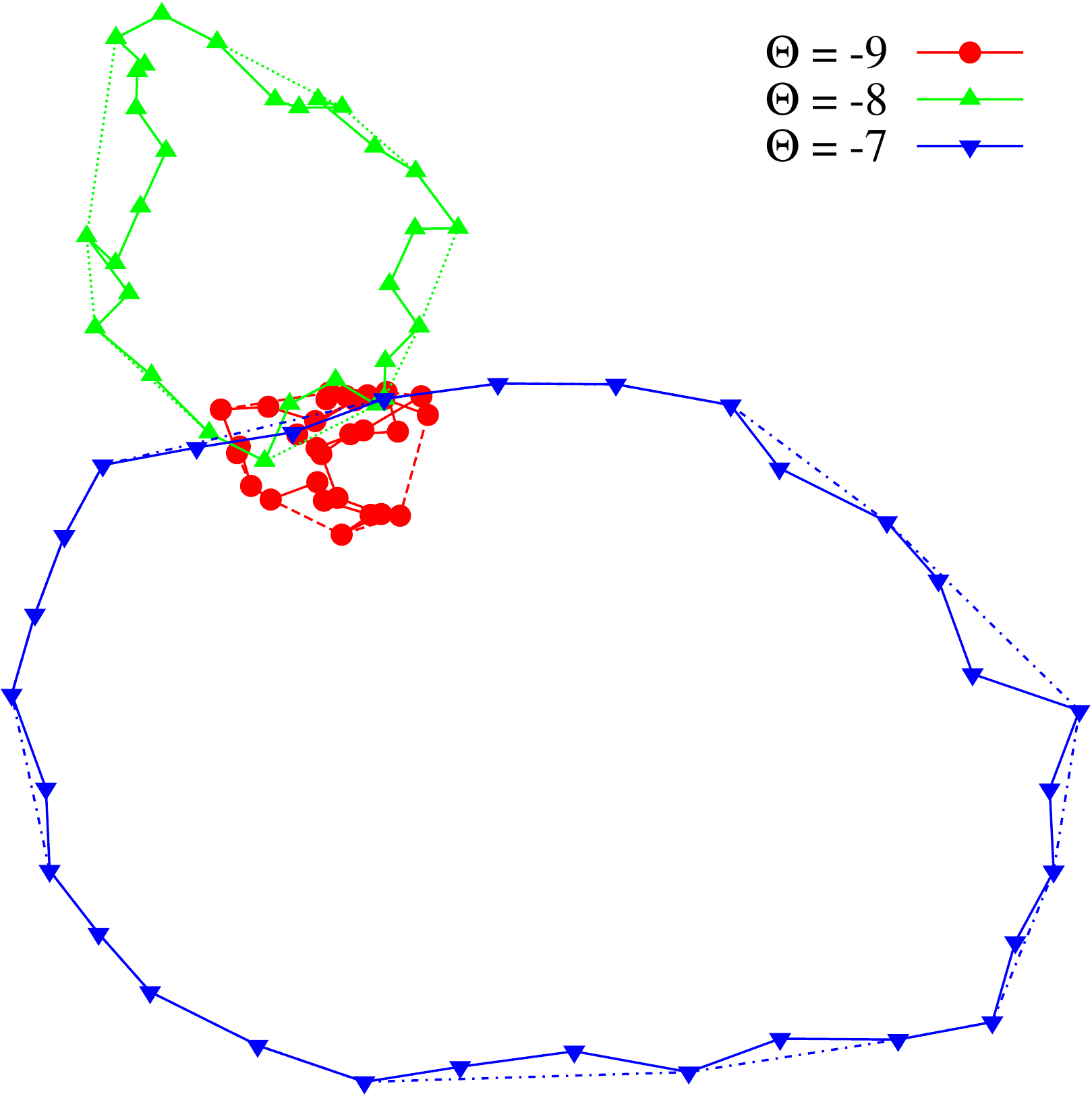}}
\caption{(color online) 
Different renditions of closed walks of length $T=30$ after 
$T_{\text{MC}} = 3 \cdot 10^5$ Monte Carlo steps, i.e., $T_s = 10^4$ 
sweeps, at the given temperatures. The top row features lattice 
steps, while the bottom row shows Gaussian steps, plus the according 
convex hulls (indicated by the dashed lines). The walks to the left 
are open and those to the right closed. Higher temperatures inhibit
 the acceptance of step alterations which lower the hull area $A$, thus 
the walks and their hulls experience drastic growth.}
\label{fig:walks_t0_cl}
\end{figure}

The MCMC consist of an evolution of random walks $\mathcal{W}(t)$,
$t$ being another discrete ``time'' parameter, not to be confused
with the time parameter $\tau$ of the random walks.
For the walks, we measure the property $S(t)$, i.e., the area ($S=A$) or 
perimeter ($S=L$), depending on which distribution we are aiming at.
The initial configuration $\mathcal{W}(0)$ is any walk configuration,
e.g., a randomly chosen one.
At each Monte Carlo step $t$, the walk $\mathcal{W}(t)$ is altered to 
$\mathcal{W}^*$ by replacing a randomly selected step $\vec{\delta}_i$
($i\in\{1,2,\ldots,T\}$) 
with a newly generated step $\vec{\delta}'_i$. The new step is generated
according the same distribution as all random walks steps
of the corresponding type, e.g., Gaussian.
The convex hull of 
$\mathcal{W}^*$ is calculated, leading to quantitiy $S^*$. 
The alteration $\mathcal{W}^*$ is \emph{accepted} ($\mathcal{W}(t+1) = 
\mathcal{W}^*$) according to the Metropolis probability:

\begin{equation}
 p_{\text{Met}} = \min\left[1, e^{-(S^*-S(t))/\Theta} \right]\,.
\end{equation}

Here, $\Theta$ is 
the (artificial) Monte Carlo ``temperature'', which is just
a parameter used to set the range of the sampled values. 
If the alteration is not accepted, it is \emph{rejected}, i.e., 
$\mathcal{W}(t+1) = \mathcal{W}(t)$.
Now, over the course of the 
MC steps, $A$ or $L$ develops according to the effected changes and 
$\Theta$. Examples are shown in Fig. \ref{fig:walks_t0_cl}. 
As usual, the MCMC time is measured in terms of \emph{sweeps},
i.e., number of steps per system size, which is here the walk length $T$.

\begin{figure}[ht!p]
\centerline{\includegraphics[width=1.0\linewidth]{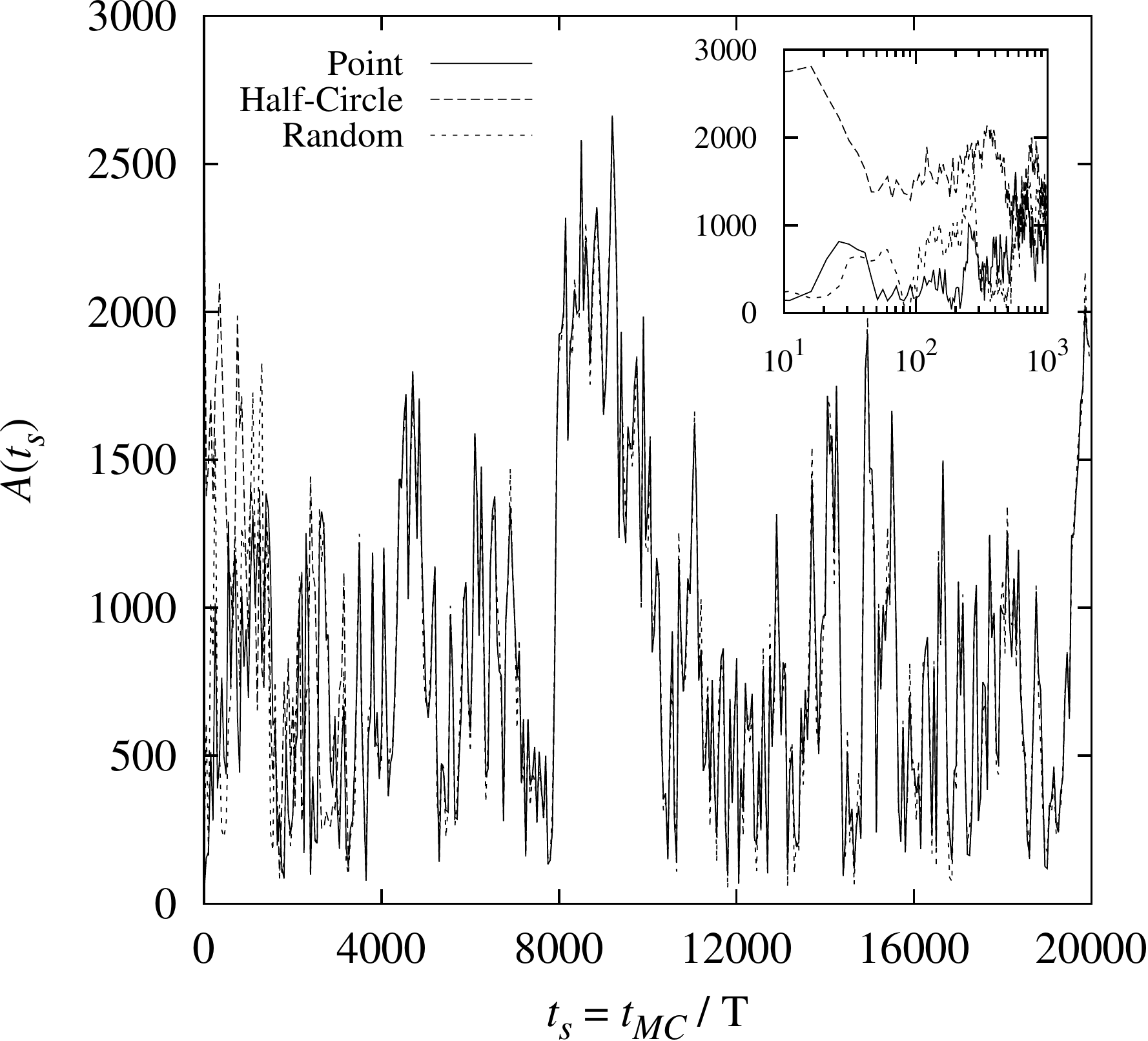}}
\caption{Equilibration of the hull area $A$ of open Gaussian
walks over the number of Monte 
Carlo steps, normalized to $t_s$ by walk length $T=50$. For 
$\Theta = -16.25$, the model is fully equilibrated after ca. 
$t_s = 5\cdot 10^4$ sweeps, and heedless of the differing initial 
configurations (see inset), i.e., $A$ behaves similarly within the
range of fluctuations.} 
\label{fig:equilibration_t0}
\end{figure}

Like in any MCMC simulation one must equilibrate the simulation, i.e.,
discart the initial part of the measured quantities until ``typical''
values are found.
Equilibration was found to be difficult but possible 
for simulations regarding hull area $A$. This was in particular difficult for 
Gaussian-distributed $\vec{\delta}_i$ components, probably due
to the fact that there exists no 
upper limit for $A$ and $L$, in contrast to the discrete case.
 As demonstrated in Fig.\ 
\ref{fig:equilibration_t0} $\Theta=-16.25$ it still 
took several $t_s \propto 10^5$ MC sweeps until equilibration. Opposed 
to this, simulations regarding perimeter $L$ equilibrated rather 
quickly, typically within less than $t_s \approx 10^3$ sweeps. 
This differing behavior seems to result from the fact that the 
replacement of one single step $\vec{\delta}_i$ may affect much 
larger changes in $A$ than in $L$, thereby making the whole model much 
more sensitive, especially for small value of $T$.

For a given walk type (discrete/Gaussian, open/closed), a given
walk length $T$ and a given quanity ($A$ or $L$) we performed
simulations for different values of $\Theta=\{\Theta_1,\ldots,\Theta_K\}$. 
The numer $K$ of temperatures and the actual values, 
posive and negatives ones, depend heavily on the model, the walk length
and the measured quantity, see below how the temperatures are chosen. 
Thus, we got
different distributions $P_{\Theta}(S)$ which are related with the 
actual distribution $P(S)$ according to the following relation
\cite{align2002}

\begin{equation}
 P(S) = e^{S/\Theta} \cdot Z(\Theta) \cdot P_{\Theta}(S)
\end{equation}

Note that ``distribution'' here either means ``probability''
or ``density'', depending on whether the possible values
of $S$ are discrete or continuous.
For different values of $\Theta$, different ranges of the
measured value $S$ were obtained. This allows for a picewise
reconstruction of $P(S)$.
It only requires knowledge of the normalization constants 
$Z(\Theta)$.  They can be calculated through inversion of this 
formula whenever for two values $\Theta_1$ and $\Theta_2$ the
ranges of the sampled values of $S$ overlap.  Thus, the temperatures
were chosen such that for neighbouring temperatures the measured
histograms sufficiently overlap.
In principle,
for each value of $S$ where the measured values $P_{\Theta_1}(S)>0$
and $P_{\Theta_2}(S)>0$ one estimate of $Z(\Theta_1)/Z(\Theta_2)$
is obtained.
Note that in case of 
equlibration, this ratio is more or less constant for all values
of $S$ where the two histograms overlap. One the other hand, 
for non-equilibrated cases a systematic dependence on $S$ is seen. In this
way we have another convenient criterion to verify equilibration.
Via these pairwise comparisons, all ratios $Z(\Theta_k)/Z(\Theta_{k+1})$
can be determined. Finally, the overall determination of the
normalization constants is obtained from the global normalization
constraint $\sum_S P(S)=1$ (or $\int P(S)\,dS=1$ for the continuous case).


\section{Results \label{sec:results}}

\subsection{Independent Points}

To verify our simulations, we first simulated convex hulls of 
$n$ independently distributed points in the unit square $[0, 1]^2$.
 For this case some mathematical results are known 
\cite{Cabo1994,Buchta2005,Groeneboom2012}. In particular the remaining 
area $\tilde{A}_n = 1 - A(n)$ outside the convex hull is considered. 
The distribution of the rescaled remaining area 
\begin{equation}
{(\tilde{A}_n - 4b_n})/2c_n\,,
\label{eq:remainin:area}
\end{equation}
 where $b_n = \frac{2}{3}\frac{\log n}{n}$ 
and $c_n = \sqrt{\frac{28}{27}\frac{\log n}{n^2}}$, should converge for 
$n \rightarrow \infty$ in distribution to the standard normal 
distribution $\mathcal{N}(0, 1)$.

\begin{figure}[ht!p]
\centerline{\includegraphics[width=1.0\linewidth]{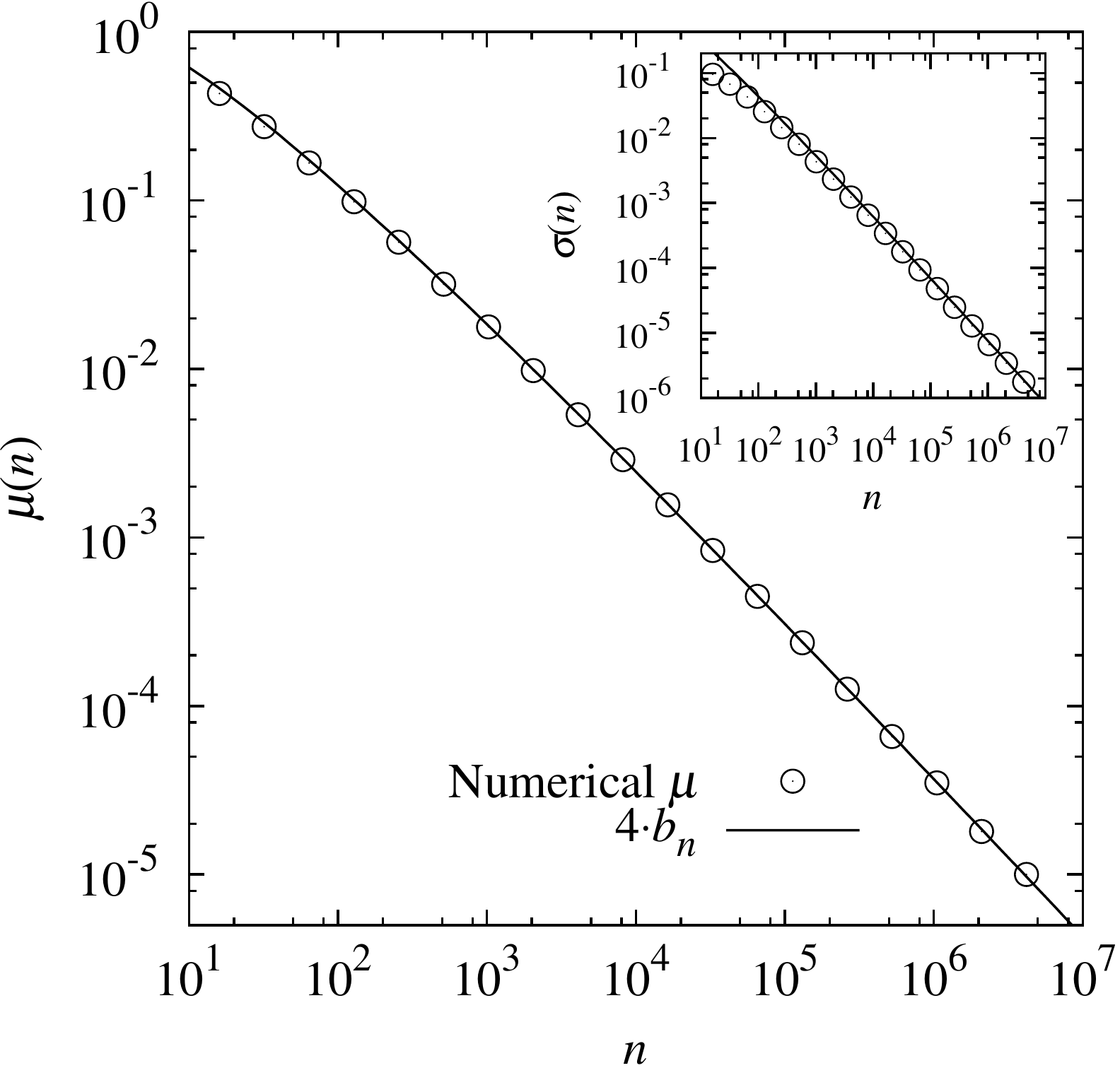}}
\caption{Average $\mu$ and variance $\sigma^2$ (inset) of the 
remaining area $\tilde{A}_n$ of a unit square in presence of a convex 
hull of $n$ randomly distributed points. The symbols show the 
numerical results while the lines 
show $4b_n$ and $2c_n$, respectively.
}
\label{fig:squares}
\end{figure}

Thus, for the unscaled data $A(n)$, the mean $\mu$ and the 
standard deviation $\sigma$ should follow $4b_n$ and $2c_n$, 
respectively. This 
is the case for our simulations, as visible in 
Fig.\ \ref{fig:squares}, where we indeed find a $4b_n$ behavior for the
mean and for larger sizes a $2c_n$ behavior for the variance.

\begin{figure}[ht!p]
\centerline{\includegraphics[width=1.0\linewidth]{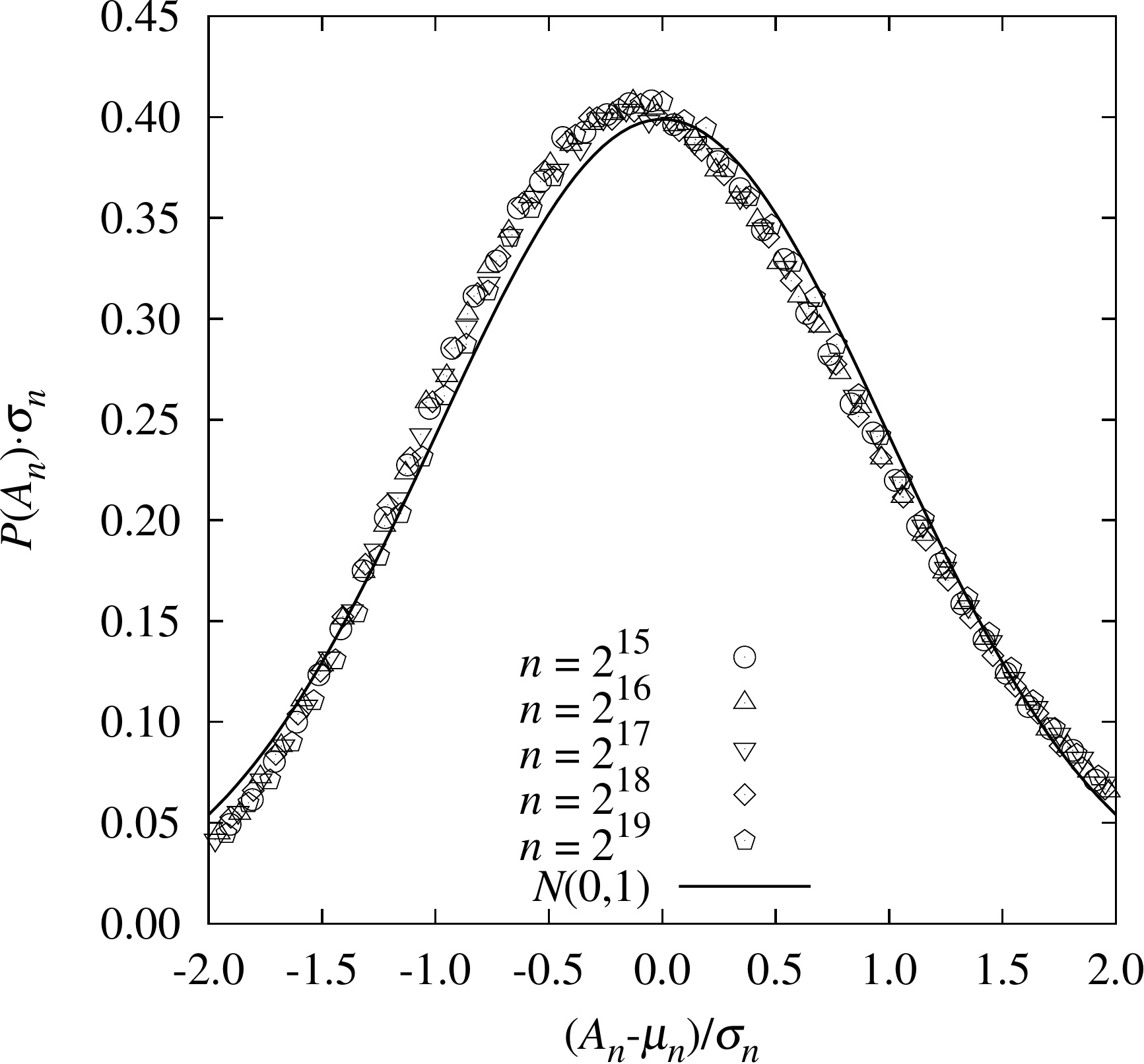}}
\caption{As predicted by Cabo \& Groeneboom \cite{Cabo1994}, 
the distributions $P(A_n)$ of the remaining area \emph{not} covered by 
the convex hull over $n$ random points inside the unit square 
should converge towards $\mathcal{N}(0, 1)$ when being rescaled according
to Eq.\ \ref{eq:remainin:area}.}
\label{fig:squares_rescaled}
\end{figure}

In a similar way, we attempted to visualize 
the predicted convergence of the distributions towards 
$\mathcal{N}(0, 1)$ by plotting $P(\tilde{A}_n)\cdot\sigma$ over 
$(\tilde{A}_n-\mu)/\sigma$, cf. Fig. \ref{fig:squares_rescaled}.
As visible, the measured density is very close
to $N(0,1)$ but no clear convergence is visible. Probably
this would be visible only for much larger number $n$ of points,
thus the convergence is very slow.
Note that these deviations become more pronounced in the small-probability
tails, which we also obtained wihthin the MCMC approach for
selected numbers $n$ (not shown here).
Anyway, for the purpose of verifying our simulations, the agreement 
with the predicitions is sufficient.

\subsection{Random Walks}

\begin{figure*}[htbp]
\centerline{
\includegraphics[width=0.324\linewidth]{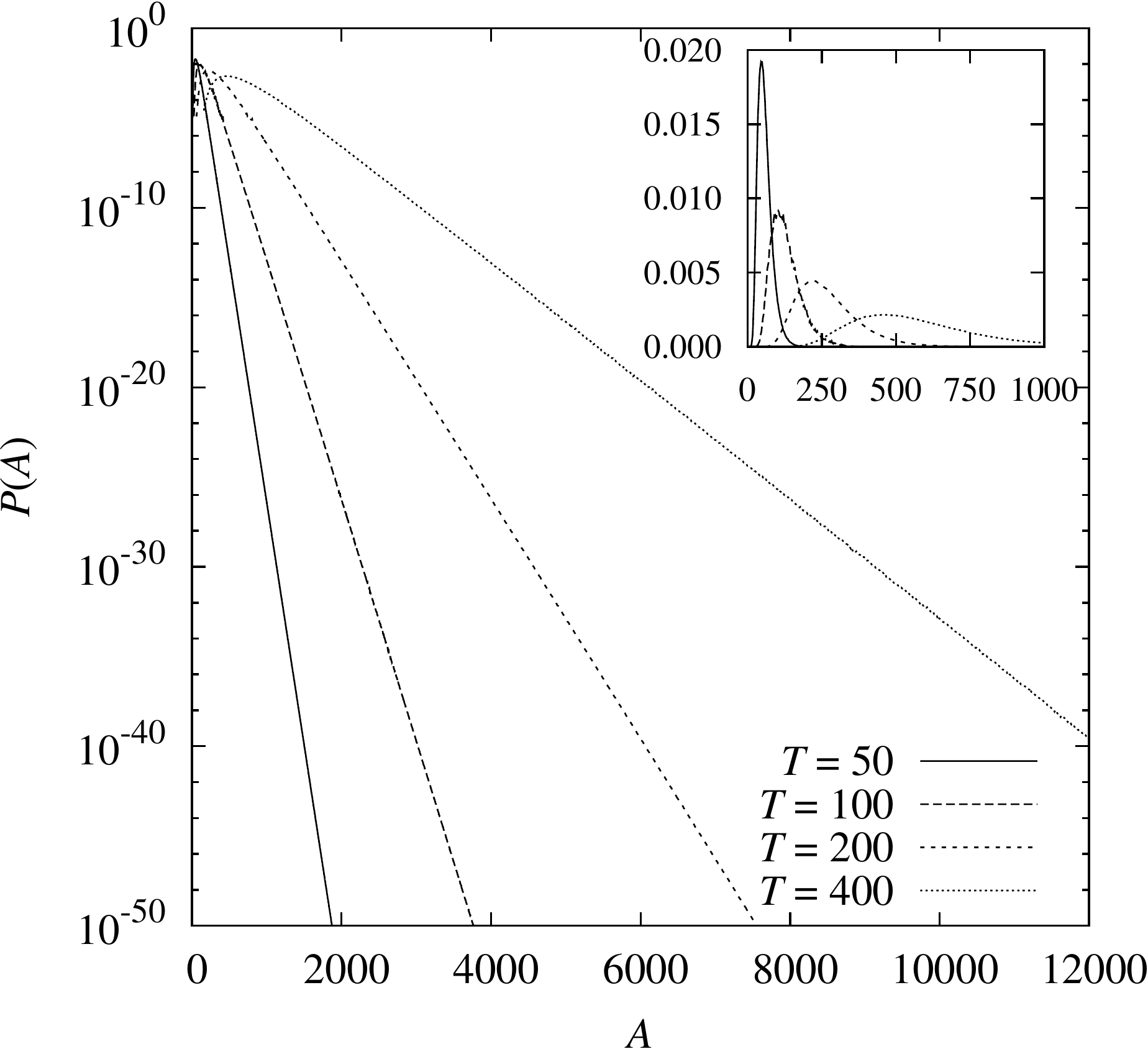} 
\includegraphics[width=0.325\linewidth]{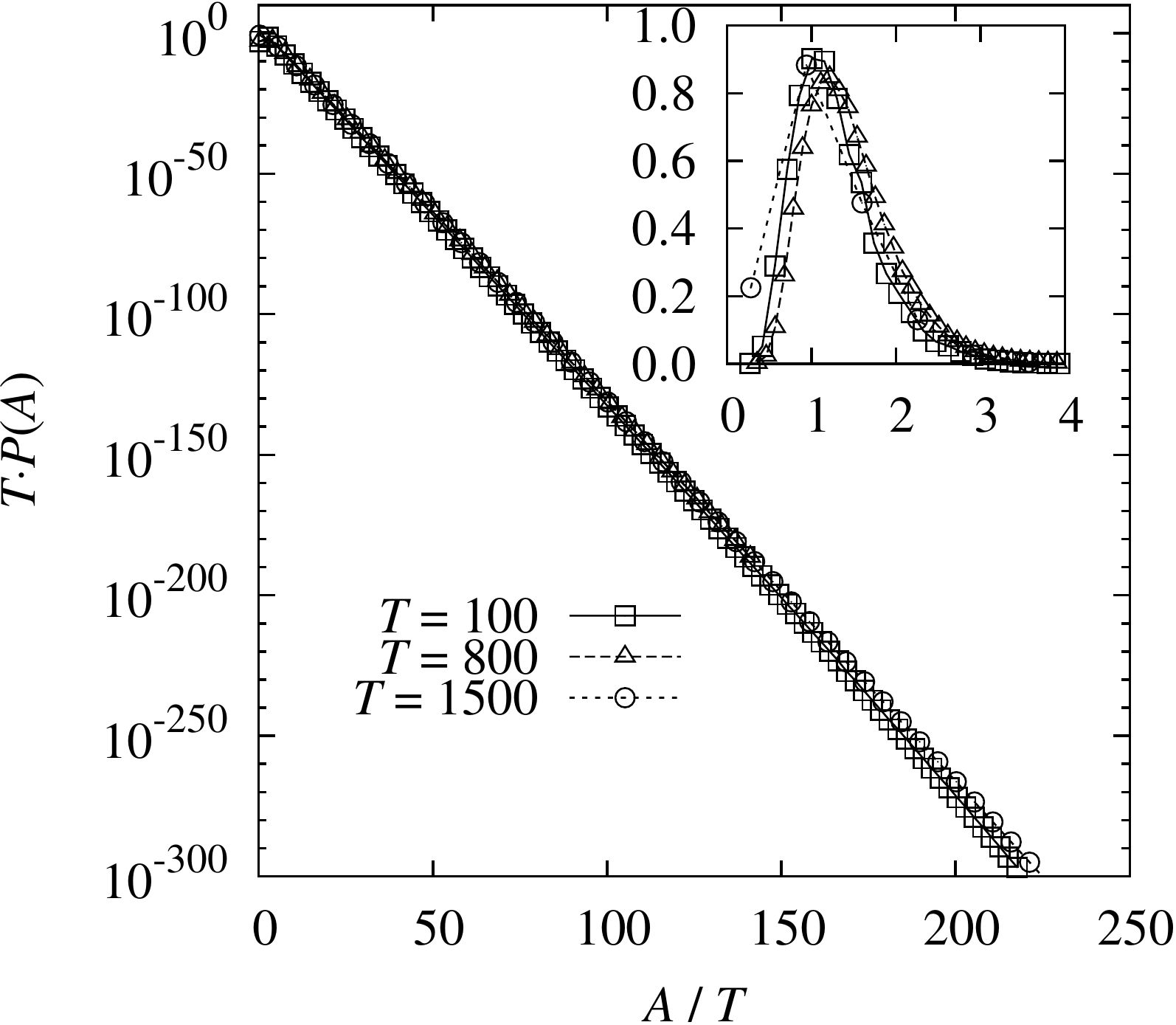} 
\includegraphics[width=0.325\linewidth]{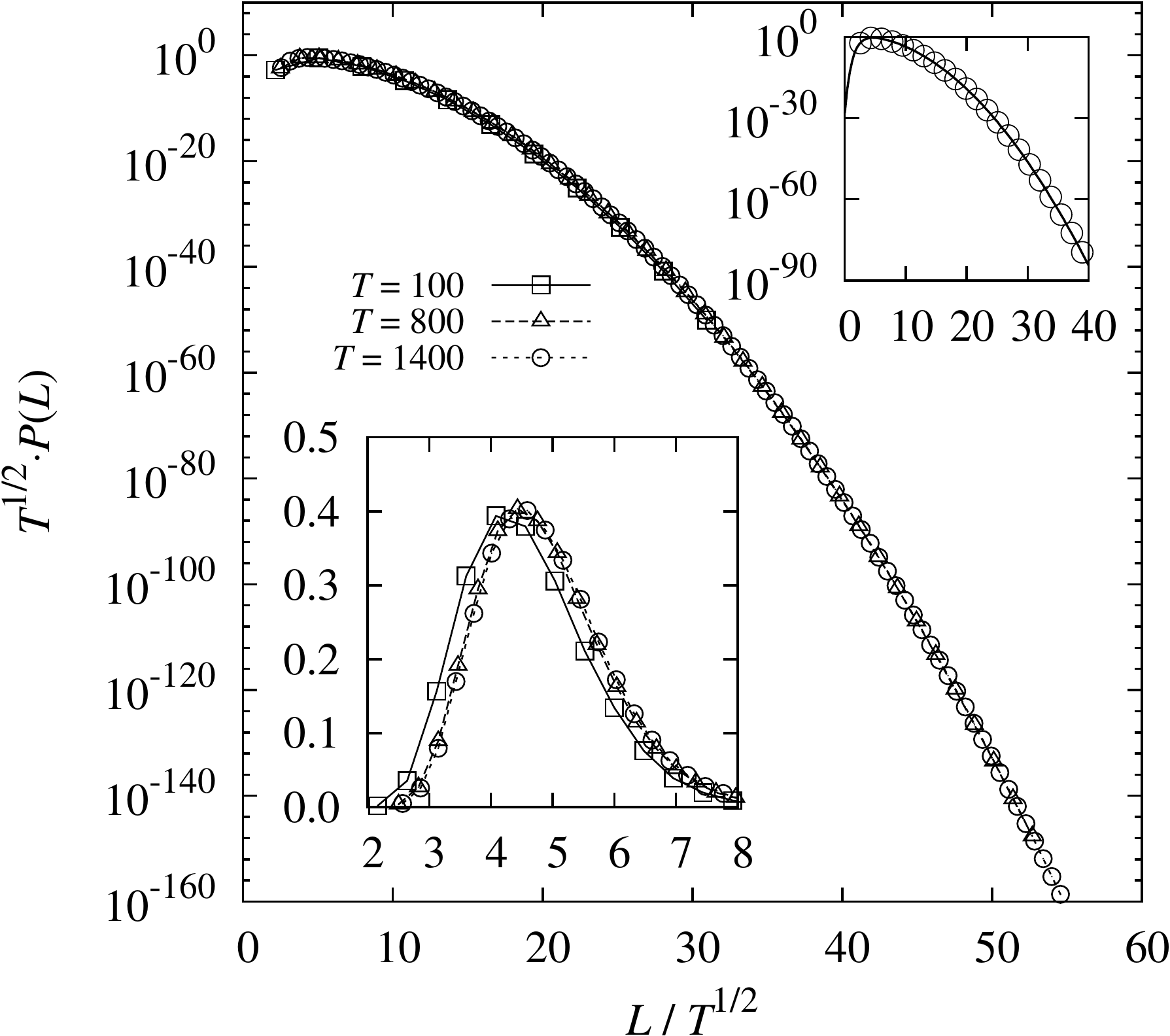} 
}
\caption{Distributions $P(A)$ and $P(L)$ of area $A$ and perimeter $L$ 
of the convex hull over Gaussian walks of different lengths $T$. 
 (Left) ``original'' distribution $P(A)$ for selected small values
of the walk length $T$. Main: plot log scale, inset: peak region
in linear scale. (Middle) rescaling of $P(A)$
according to Eq.\ \eqref{eq:scaling_area}. Main plot: log scale; upper
right inset: peak region in linear scale; lower left inset:
fit of Gumbel distribution to $T=1500$ data  (Right)  rescaling of $P(A)$
according to Eq.\ \eqref{eq:scaling_perimeter}. Main plot: log scale;
lower left inset: peak region in linear scale; upper right inset: fit of
$T=1400$ data to ``modified'' Gumbel distribution.
\label{fig:distr:P:Gaussian}}
\end{figure*}

For each type of walk and either of the two quantities $A$ and $L$,
we considered walks lengths 
$T \in [30,2000]$ with ca. $K = 10^6$ samples per temperature $\Theta$. 
Through sampling multiple values of $A$ or $L$ per MC run (i.e., each 
$t_s = 10$ sweeps), it was possible to obtain the results for each case within 
several days of CPU time on a average-size multi-core cluster (using up to few 
dozens of cores per case, corresponding to the number of temperatures
$\Theta$).

An example of the resulting distributions, $P(A)$ for the case of Gaussian
open walks of various lengths, is shown in Fig.\ \ref{fig:distr:P:Gaussian}.
Evidently, we were able to obtain the distributions over a large range of the
support, with probability densities as small as $10^{-300}$.

As the systematic behavior of the distributions $P(A)$ for different 
walk lengths $T$ hints a common origin at a distribution 
$\tilde{P}(A)$ (cf. Fig. \ref{fig:distr:P:Gaussian}), the first property 
we put under scrutiny  
is the scaling behavior. The scaling behavior \cite{Majumdar2010} of the mean 
$\langle A \rangle \sim T$
suggests that distributions $P(A)$, which in fact depends on the walk length $T$,
 scales according to the  following relation:

\begin{equation}
 P(A) = \frac{1}{T} \cdot \widetilde{P}\left(\frac{A}{T}\right)\,, 
 \label{eq:scaling_area}
\end{equation}
where $\widetilde{P}$ is a $T$-independent distribution.
 The application of the scaling assumption leads 
to a reliable collapse of the distributions $P(A)$ towards 
$\widetilde{P}\left(\frac{A}{T}\right)$ with the exception of 
finite-size effects within the tail, as shown in 
the middle of Fig.\ \ref{fig:distr:P:Gaussian}.

For perimeter distributions, we use a similar scaling assumption, which 
is based upon $\langle L \rangle \propto \sqrt{T}$ \cite{Majumdar2010}:

\begin{equation}
 P(L) = \frac{1}{\sqrt{T}} \cdot \widetilde{P}\left(\frac{L}{\sqrt{T}}\right) 
\label{eq:scaling_perimeter}
\end{equation}

The application of this relation to open walks, shown in Fig.\ 
 \ref{fig:distr:P:Gaussian}, reveals an almost perfect collapse 
even for large values of $T$ and the latter parts of the rare-event tail. 
We found similar scaling (not shown) for the discrete cases. 
Opposed to this, closed walks seem to be more strongly affected 
by finite-size effects, i.e., to tend towards 
$\widetilde{P}\left(\frac{L}{\sqrt{T}}\right)$ rather slowly. Nevertheless,
in summary for all 8 cases the corresponding scaling is supported by our data.

Next, to extrapolate the leading behavior of the
distribution in the large-deviation tail, 
we calculate the empirical rate function 
$\Phi(s)$ \cite{touchette2009} and guard its behavior over walk length 
$T$. Under the assumtition that away from the maximum
the the leading behavior of the probability is exponentially small in $T$, i.e., 
the rate function is defined as:
\begin{equation}
 \Phi(s) \equiv -\frac{1}{T} \log P(s)\,.
\label{eq:rate:function}
\end{equation}
Distributions where the rate function is well defined, i.e., follow
Eq. \eqref{eq:rate:function} are said to
obey a ``large-deviation principle''.
To allow  a comparison and extrapolation of the rate function, 
$s$ is usually a quantity normalized with the maximum
possible value, such that $s\in [0,1]$. Thus,  
one would define $s_A = \frac{A}{A_{\max}}$ and 
$s_L = \frac{L}{L_{\max}}$, accordingly. For the lattice walk model 
with $J = \sqrt{2}$, these maximum possible values were found to be 
$A_{\max} = \frac{T^2}{4}$  and $L_{\max} = 2\cdot\sqrt{2}\cdot T$ for 
open walks as well as  $A_{\max} = \frac{T^2}{8}$  and 
$L_{\max} = \sqrt{2}\cdot T$ for closed walks. But for Gaussian walks 
no actual ``maximum'' values of $A$ and $L$ exist, so in remembrance of 
the lattice cases to obtain the same scaling, we choose $s_A = \frac{A}{T^2}$ or 
$s_L = \frac{L}{T}$, respectively. 

\begin{figure}[htbp]
\centerline{\includegraphics[width=0.8\linewidth]{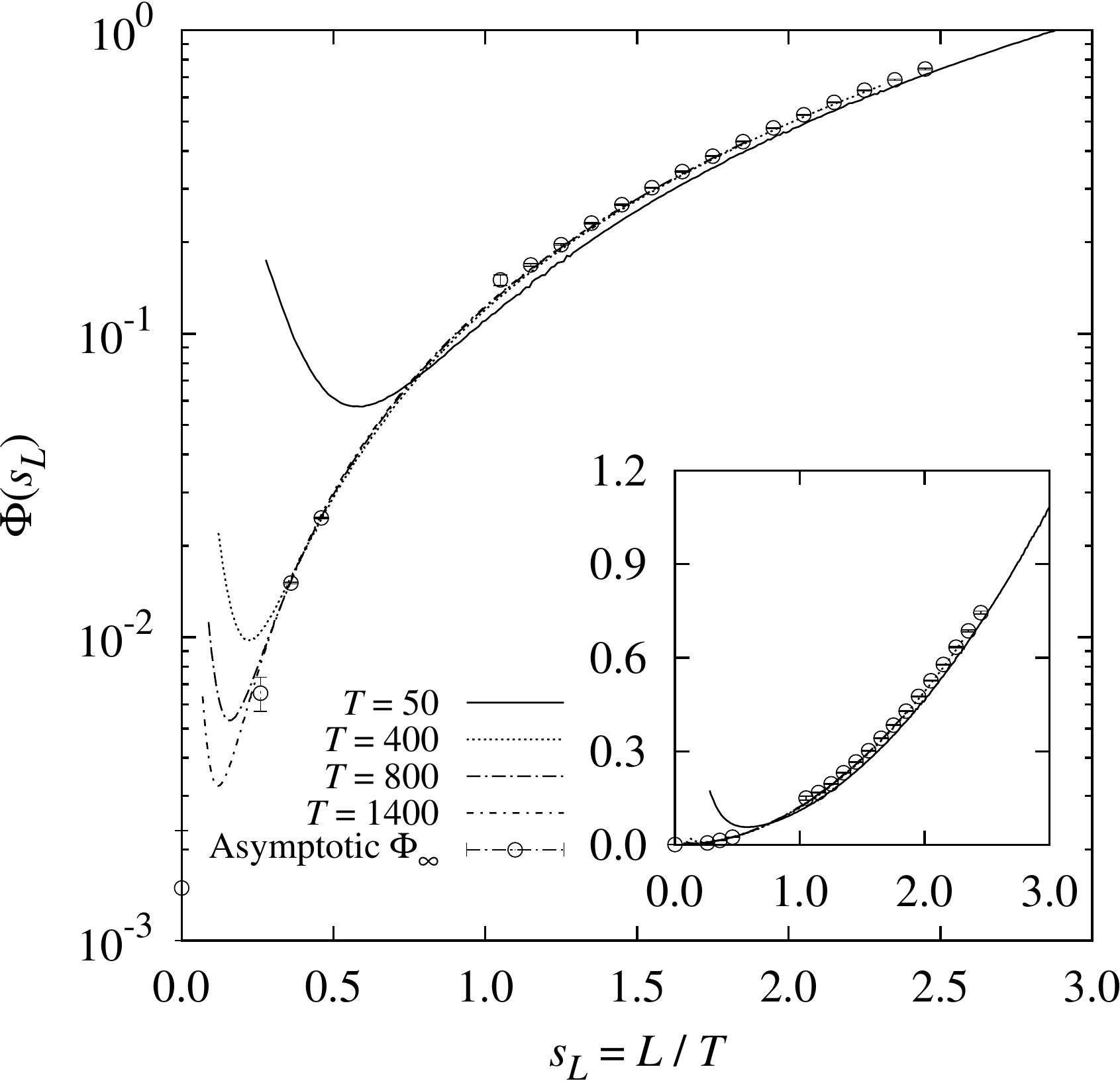}}
\caption{The rate functions $\Phi(s_L)$ 
of open Gaussian walks for different walk lengths (lines). The symbols show
the extrapolated values ($\Phi_{\infty}(s)$ see text and 
Fig.\ \ref{fig:p1a_rf_t0_op_peri_asymptotic}). 
The inset shows the same on normal scale.
\label{fig:p1a_rf_t0_peri_op}}
\end{figure}

\begin{figure}[htbp]
\centerline{
\includegraphics[width=0.86\linewidth]{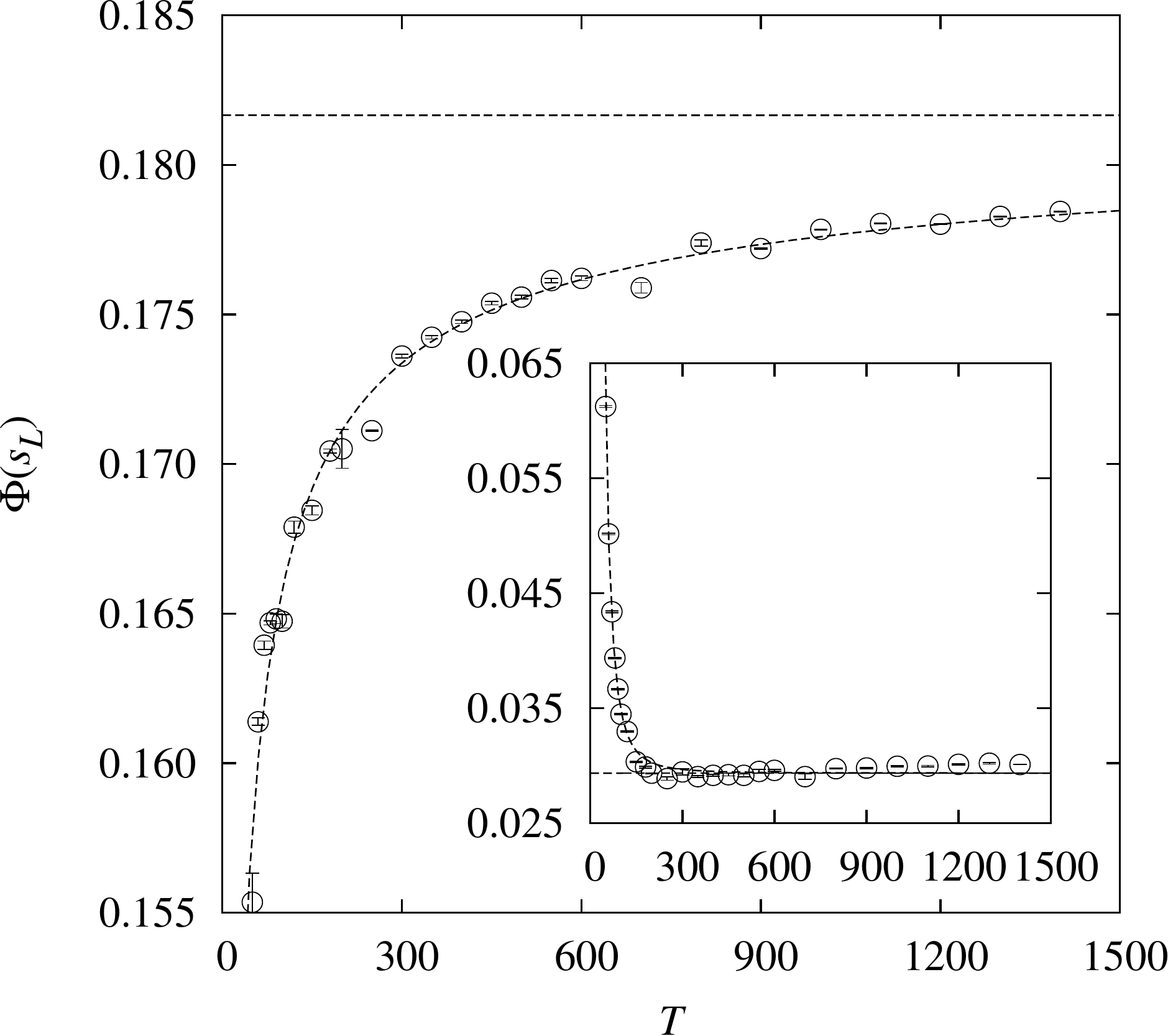} }
\caption{Dependence of rate function $\Phi(s_L)$ (for open Gaussian walks)
on walk length $T$ at selected
values $s = 1.2$ and $s = 0.5$ (inset). A power law was used
to determined the $T\to\infty$ convergence towards an asymptotic value 
$\Phi_{\infty}$, indicated by horizontal lines. 
\label{fig:p1a_rf_t0_op_peri_asymptotic}
}
\end{figure}

An example of the resulting rate functions, for the case of the perimeter
of the hull of open Gaussian walks, is shown in 
Fig.\ \ref{fig:p1a_rf_t0_peri_op}.
Apparently, for large values of $s_L$, the rate functions for different
walks lengths agree very well, while for small values of $s_L$
strong finite length effects are visible. Anyway, from the visual
impression, the rate function seems to converge for $T\to\infty$ pointwise 
to a limiting function, which starts at the origin. Thus, its seems that
the distribution obey the large-deviation principle.
To quantify this, we performed fits of the form
\begin{equation}
\Phi(s,T) = \Phi_{\infty}(s) - \xi \cdot T^{-\gamma}
\label{eq:power:law:fit}
\end{equation}
for many
values of $s$, as it is demonstrated 
in Fig. \ref{fig:p1a_rf_t0_op_peri_asymptotic}. 
We obtained similar results for all other walk types and measured
quantities. We found that
$\Phi_{\infty}(s)$ appears to grow according to power laws $s^{\kappa}$.
(The parameters 
$\xi$ and $\gamma$ are not of interest and show no systematic behavior with $s$).
 The results are listed in 
Tab.\ \ref{Tab:rf_kappa}. The values for the area are very close to $\kappa=1$
and for the perimeter very close to $\kappa=2$, in particular
for the Gaussian walks.

\begin{table}[htbp]
\centering
\begin{tabular}{|c||c|c|}
 \hline
 Model            & $\kappa_A$, $P(A)$ & $\kappa_L$, $P(L)$ \\\hline\hline
 Gaussian, open   & 0.999(1)         & 2.03(2)          \\\hline
 Gaussian, closed & 1.06(2)          & 2.06(1)          \\\hline
 lattice, open    & 1.18(1) &          2.13(1) \\\hline
 lattice, closed  & 1.12(2)          & 2.12(1) \\\hline
\end{tabular}
\caption{Resulting exponents $\kappa$ which deterine the power-law growth of the asymptotic value $\Phi_{\infty}(s)$ of the rate function $\Phi(s)$ with $s_A$ or $s_L$, respectively.}
\label{Tab:rf_kappa}
\end{table}

This can be already seen when one combines the rate function for the scaled
variables with the scaling forms Eqs. \eqref{eq:scaling_area}
and \eqref{eq:scaling_perimeter}. By equating the two expressions
for the distribution of the area we obtain

\begin{equation*}
e^{-\Phi(A/T^2)\cdot T} \sim \frac 1 T \widetilde P(A/T)\,.
\end{equation*}
To make the argument of the exponential a function of $A/T$ (the factor $1/T$
is of lower order and can be ignored) $\Phi(s)\sim s$ 
must hold, i.e., $\kappa_A=1$.
Correspondingly for the perimeter we get
\begin{equation*}
e^{-\Phi(L/T)\cdot T} \sim \frac {1}{\sqrt{T}} \widetilde P(L/\sqrt{T})\,,
\end{equation*}
which results in $\phi(s)\sim x^2$, i.e., indeed $\widetilde P(L/\sqrt{T})\sim$ 
$e^{-(L^2/T^2)T} =e^{-(L/\sqrt{T})^2}$, and $\kappa_L=2$.

The fact that for Gaussian walks, the exponents $\kappa$ agree very well with 
$\kappa_A=1$ and $\kappa_L=2$ but not quite for the lattice walks is probably due
to the discrete structure of the lattice and due to the limited
accessible area. These have a strong influence for the given limited
length $T$ of the walks and influence the rate function in particular for
small and large values of $s$. In fact the rate functions
for the lattice case are continuously
bending up in a log-log plot (not shown), 
such that one easily can find a short interval where the expected exponent
$\kappa$ appears. Thus, to conclude this part, the data is well described
by the rate function, i.e., the distribution obeys the above large-deviation
principle. Hence, in connection with the fulfilled scaling 
forms Eqs. \eqref{eq:scaling_area}
and \eqref{eq:scaling_perimeter}, we see that to leading order and 
asymtotically the distributions are given by $P(A)\sim e^{-A/T}$
and $P(L)\sim e^{-L^2/T}$.

\begin{figure}[ht!p]
\centerline{\includegraphics[width=1.0\linewidth]{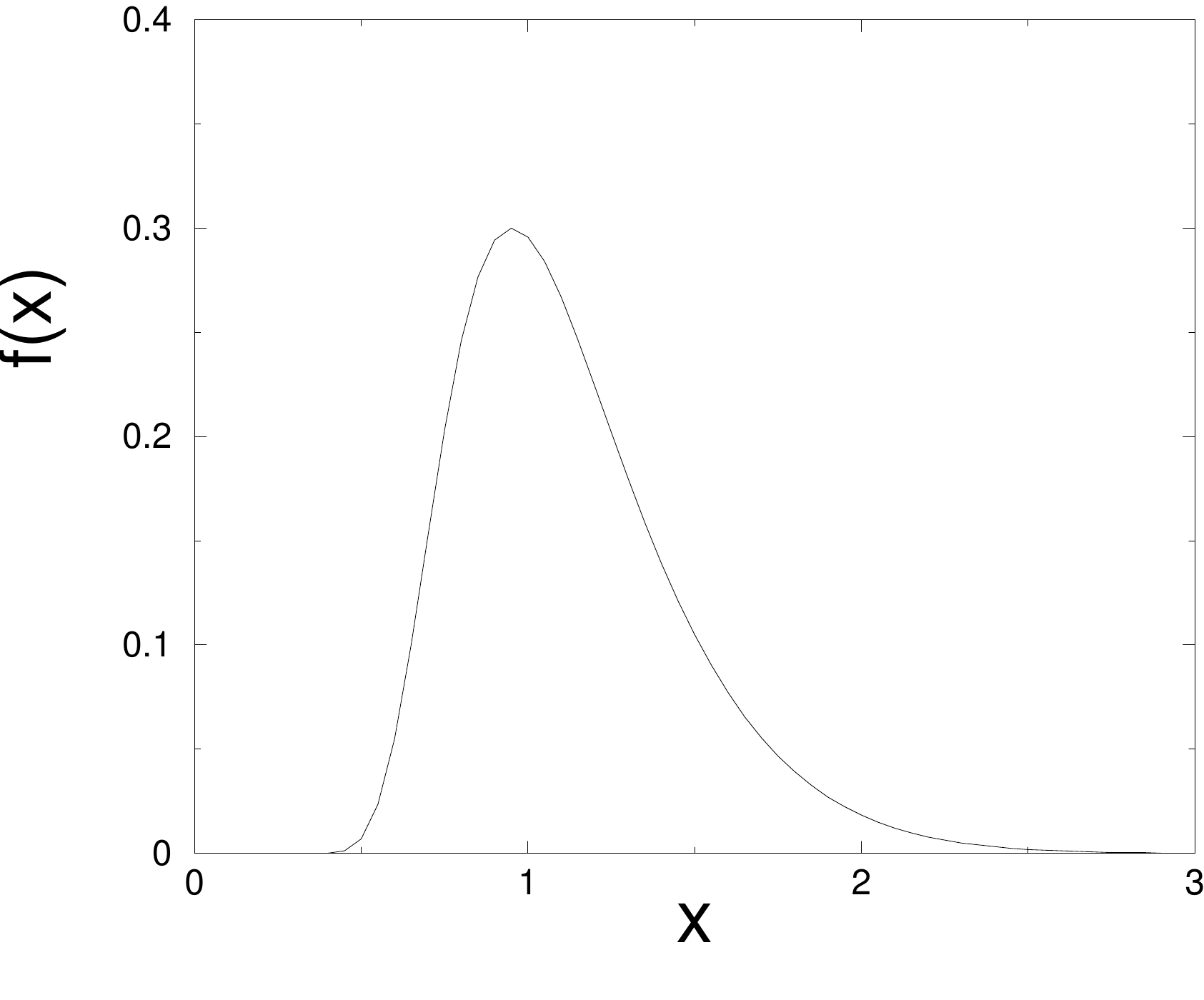}}
\caption{The scaling function $f(x)$ in Eq. (\ref{scaling_span.2})
for the span distribution for a $1$-d Brownian motion, plotted
as a function of $x$.}  
\label{fig:span}
\end{figure}

The Gaussian and the exponential
tails, respectively for the distributions of the perimeter and the area, 
may be guessed using very simple heuristic arguments. Indeed, 
the perimeter $L$ of the convex hull of a two dimensional
stochastic process morally scales as 
$s$, where $s$ represents
the span of the one dimensional component process. For the
$2$-d Brownian motion, $s$ is simply the span of a one dimensional
Brownian motion of duration $T$. The probability distribution
$P(s,T)$ of the $1$-d Brownian motion of duration $T$ and
diffusion constant $D$ is well 
known\cite{hughes1996,Kundu2013} and has a scaling form
\begin{equation}
P(s,T) = \frac{1}{\sqrt{4DT}}\, f\left(\frac{s}{\sqrt{4DT}}\right)
\label{scaling_span.1}
\end{equation}
where the scaling function $f(x)$ is given exactly by
\begin{equation}
f(x)= \frac{8}{\sqrt{\pi}}\, \sum_{m=1}^{\infty}
(-1)^{m+1}\,‚
m^2\, e^{-m^2\, x^2} \, .
\label{scaling_span.2}
\end{equation}
The scaling function has the following asymptotic behavior
\begin{eqnarray}
f(x)\to \left\{\begin{array}{ll}
2\, \pi^2\, x^{-5}\, e^{-\pi^2/{4\,x^2}}\;\;
&{x\to 0}\label{small_x}\\
& \\
\frac{8}{\sqrt{\pi}}\, e^{-x^2} \label{large_x}\;\;
&{x\to \infty}
\end{array}
\right.
\end{eqnarray}
and is plotted in Fig.\ref{fig:span}.

Thus, for large $s$, $P(s,T)\sim \exp\left[-s^2/{4DT}\right]$ has a
Gaussian tail. Consequently since $L\sim s$, one would expect
a Gaussian tail for the distribution $P(L)$. Note that for small
$S$ (and hence for small $L$), one would expect an essential
singular behavior $P(L)\sim \exp[-b/L^2]$ from Eq. (\ref{small_x}),
where $a$ is some constant. The scaling function $f(x)$ in Eq. 
(\ref{scaling_span.2}), and hence $P(L)$ is almost flat near $L=0$, and 
falls off
as a Gaussian for large $L$ with a peak in between, around which 
is has an approximate Gaussian shape (see Fig. \ref{fig:span})
\begin{equation}
P(L) = \frac{1}{\sqrt{2 \pi \sigma^2}} \cdot e^{-\frac{(L-\mu)^2}{2 
\sigma^2}}
 \label{eq:Gaussian}
\end{equation}

Subsequently, for the area we can approximately assume 
$A \propto L^2$, and thus the distribution of $\sqrt{A}$ should also 
be Gaussian, including a factor $1/\sqrt{A}$ originating from
$dL/dA \sim 1/\sqrt{A}$ :

\begin{equation}
P(A) = \frac{1}{\sqrt{2 \pi \sigma_A^2 \cdot A}} \cdot 
e^{-\frac{(\sqrt{A}-\mu_A)^2}{2 \sigma_A^2}}
 \label{eq:Gaussian_area}
\end{equation}

\begin{figure}[htbp]
\centerline{\includegraphics[width=0.8\linewidth]{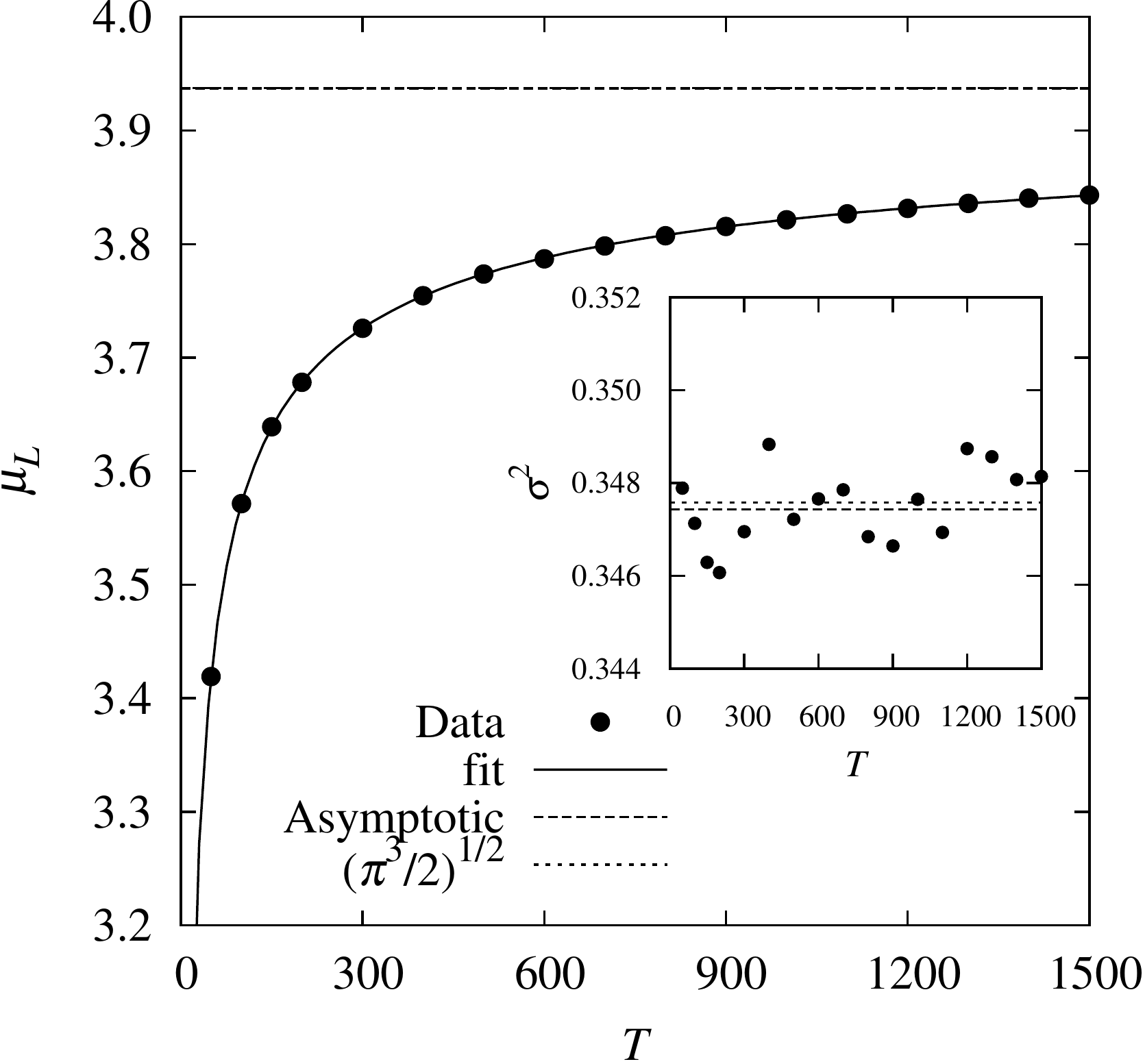} }
\caption{Average (rescaled) perimeter $\mu$ and variance $\sigma^2$
for closed Gaussian as a function of the walk length $T$. 
A convergence to the analytical expectations $\mu \approx 3.937$ and 
$\sigma^2 \approx 0.348$ is visible.
\label{fig:p1a_average_t0_cl_peri}}
\end{figure}

Had the assumption of a Gaussian distribution for the
perimeter been entirely correct, it would be sufficient to take
the average and the variance and plot the resulting Gaussian 
together with the data. First, we concentrate on the case of the hull
perimeter for closed planar Brownian motions, where mean and variance are
available analytically~\cite{Goldman1996}. The mean scaled perimeter is
given by~\cite{Goldman1996}
\begin{equation}
\mu \equiv E\left(\frac{L}{\sqrt{T}}\right) = \sqrt{\frac{\pi^3 T}{2}}\cdot \frac{1}{\sqrt{T}} = 
\sqrt{\frac{\pi^3}{2}} \approx 3.937
\end{equation}
where $E$ stands for the expectation value.
For the variance, we use Goldman's result for the second moment~\cite{Goldman1996}, 
\begin{equation}
 E((L/\sqrt{T})^2) = \frac{\pi^2}{3} \left(\pi \int^{\pi}_0 \frac{\sin u}{u} 
\text{d}u - 1\right)
\end{equation}
The value of the integral of $\int_0^{\pi}du \sin(u)/u$ is approximately 1.852,  
known as the Wilbraham-Gibbs constant. This then leads to the variance:
\begin{eqnarray*}
 \sigma^2 & = & E((L/\sqrt{T})^2)-E(L/\sqrt{T})^2 \\
& = & \frac{\pi^3}{3}(\pi\cdot1.852 - 1)-\frac{\pi^3}{2} \approx 0.348
\end{eqnarray*}
Before comparing the actual distribution, we compare average and variance
obtained from
our numerical results. By fitting power laws plus a constant, similar
to \eqref{eq:power:law:fit}, we found $\mu$ to converge asymptotically towards 
$\mu_{\infty} = 3.937(1)$, while $\sigma^2$ fluctuated slightly around 
an average of $\sigma^2_{\infty} = 0.347(1)$, see 
Fig.\ \ref{fig:p1a_average_t0_cl_peri}. Both values agree with 
the aforementioned predictions. The results of the extrapolations
for all considered cases of 
the different walk models are listed in Tab. \ref{Tab:Gauss_results}, 
and it can be seen that the asymptotic values of the parameters of 
the distributions agree rather well 
between lattice and Gaussian walks, as expected.
  Note that for the fitting the results the
stated error bars are only of statistical nature, thus do not include 
systematic
contributions due to the unkown scaling behavior. Hence, the real
error bars should be considerably larger.
Thus the results for the averages $\mu_{\infty}$ can be considered  
to match well the corresponding calculations in 
Refs.\cite{Randon-Furling2009,Majumdar2010} which read:

\begin{align}
 \langle A_{\text{op}} \rangle = \frac{\pi T}{2} \Leftrightarrow \mu_{\infty} = \frac{\pi}{2} \approx 1.571\\
 \langle A_{\text{cl}} \rangle = \frac{\pi T}{3} \Leftrightarrow \mu_{\infty} = \frac{\pi}{3} \approx 1.047 \\
 \langle L_{\text{op}} \rangle = \sqrt{8 \pi T} \Leftrightarrow \mu_{\infty} = \sqrt{8 \pi} \approx 5.013
\end{align}

\begin{table}[htbp]
\centering
\centerline{
\begin{tabular}{|c||c|c||c|c|}
 \hline
 Model                & Gaussian & Gaussian & lattice  & lattice \\
                      & open     & closed   & open     & closed  \\\hline\hline
 \multicolumn{5}{|c|}{Area distributions $P(A)$}  \\\hline\hline
 $\mu_{\infty}$       & 1.585(1) & 1.049(1) & 1.577(1) & 1.049(1) \\\hline
 $\sigma^2_{\infty}$  & 0.312(1) & 0.088(1) & 0.309(1) & 0.089(1) \\\hline\hline
 \multicolumn{5}{|c|}{Perimeter distributions $P(L)$}  \\\hline\hline
 $\mu_{\infty}$       & 5.023(2) & 3.937(1) & 5.015(1) & 3.938(1) \\\hline
 $\sigma^2_{\infty}$  & 1.075(1) & 0.347(1) & 1.075(1) & 0.348(1) \\\hline
\end{tabular}
}
\caption{Resulting asymptotic values of average $\mu$ and variance 
$\sigma^2$ of rescaled hull area $A$ or perimeter $L$ of the mentioned walk 
models. These values have been obtained by fitting to the results of 
$\mu(T)$ and $\sigma^2(T)$ with growing walk length, hereby 
extrapolating the $T\rightarrow\infty$ case. For equal walk 
topologies (open/closed walks) and quantities (area/perimeter), 
the asymptotic values agree heedless of the actual step type, i.e., the 
walk geometry.}
\label{Tab:Gauss_results}
\end{table}

However, as demonstrated in Fig.\ \ref{fig:distr:P:Gaussian}, when actually
plotting the numerical data together with a Gaussian parametrized
by the analytic values, one sees that distribution only 
matches the numerics in  the main region of the measured distributions $P(L)$, 
while the  tail exhibits a different behaviour. On the other
hand it is indeed possible to fit 
Eq.\ \eqref{eq:Gaussian} to the tail of $P(L)$ with good precision.
This means the shape of the distribution is Gaussian there. Nevertheless,
the resulting value of 
$\sigma^2\approx 1.35$ is considerably larger than the aforementioned, and
 the corresponding value of $\mu\approx 0.1$ for these fits seems useless. 
In the same way, for the other cases of perimeter distributions (not shown),
one can fit Gaussians well either to the main region of the data or to the
tails. But it is not possible to fit the
data using one single Gaussian over the full support.
Thus, the approximations which lead to the assumption of a
 Gaussian distribution were
a bit too strong and the true distributions seem to be very Gaussian-like, but
slightly different.

\begin{figure}[htbp]
\centerline{
\includegraphics[width=0.8\linewidth]{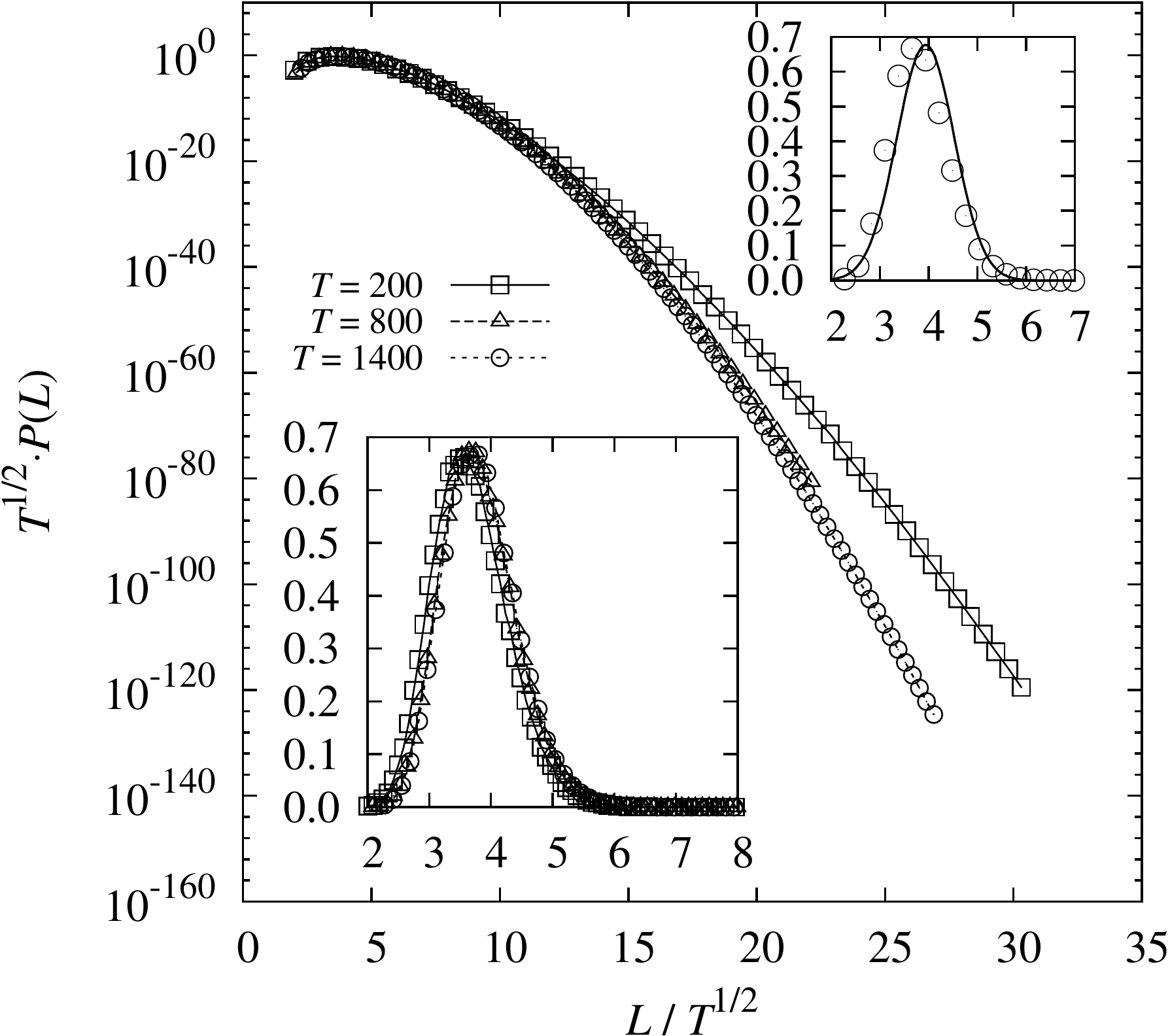} }
\caption{Scaled distributions of the 
perimeter $L$. While the scaling assumption collapses the distributions 
$\tilde{P}(L)$ onto similar behavior within the tail as well as in the 
main region (large inset), the Gaussian distribution according to 
Eq.\ \eqref{eq:Gaussian} (small inset) with the values of 
$\mu$ and $\sigma^2$ taken from
the analytics suits only in the main region. 
\label{fig:p1a_scaled_t0_peri_cl}}
\end{figure}

\begin{figure}[htbp]
\centerline{
\includegraphics[width=0.8\linewidth]{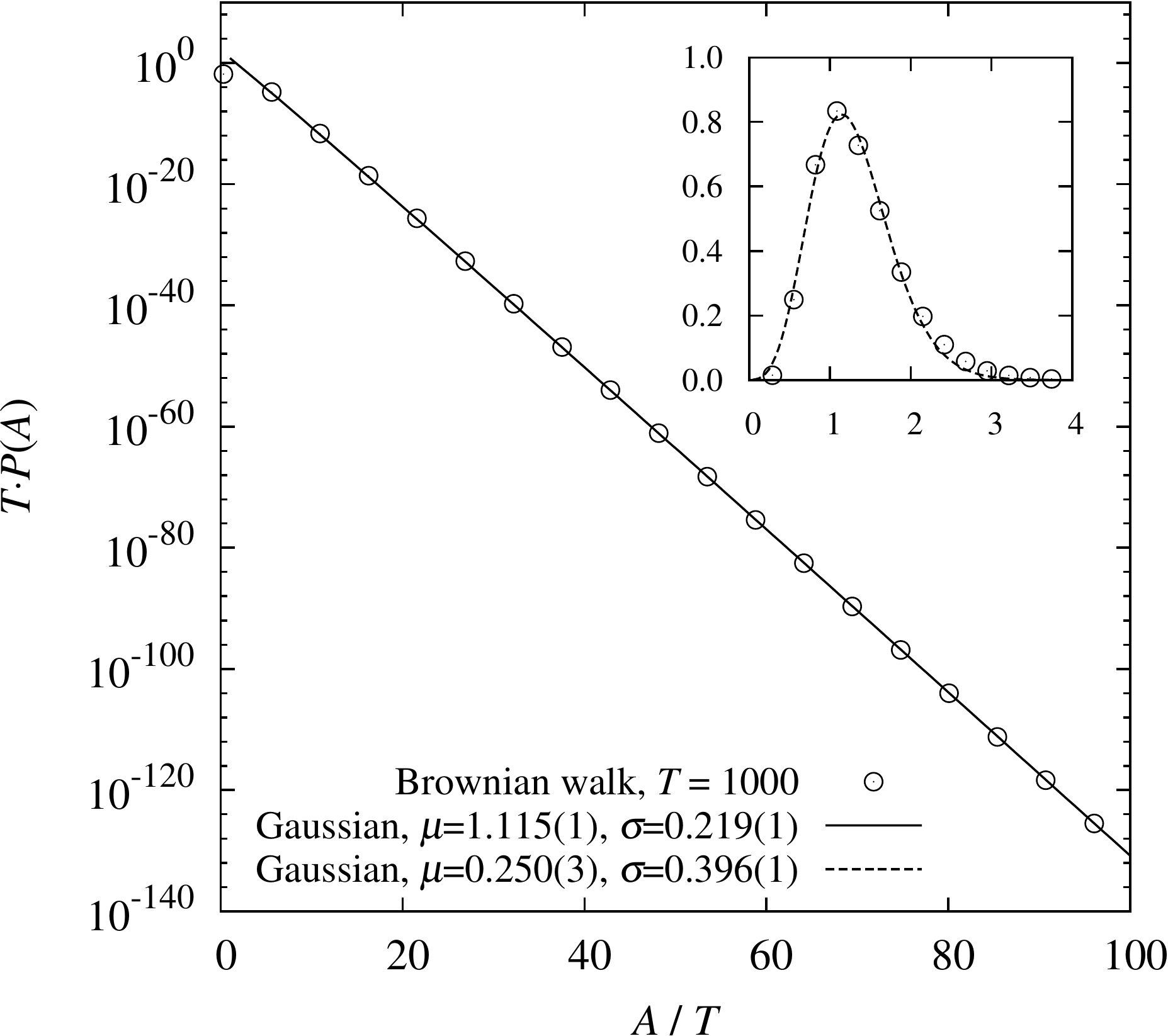} }
\caption{ Fit of a Gaussian according to 
Eq. \eqref{eq:Gaussian_area} to $P(A)$ with a focus on either 
the tail (large plot) or the region around the average (inset).
It was not possible to fit one single Gaussian over the full support.
\label{fig:p1a_gaussian_t0_area_op}}
\end{figure}

Regarding the area distribution $P(A)$, the mentioned propierties 
are demonstrated in Fig.\ \ref{fig:p1a_gaussian_t0_area_op}. Given the
previous results for the perimeter, we here just performed fits in
a similar way. A Gaussian 
of the form of Eq. \eqref{eq:Gaussian_area}  can indeed be employed 
to fit the area distribution either in the high-probability region
or everywhere away from the central region of the distribution, but again
not over the full support. Thus, the same concusion as for the perimeter
distribution holds.

\begin{figure}[htbp]
\centerline{
\includegraphics[width=0.8\linewidth]{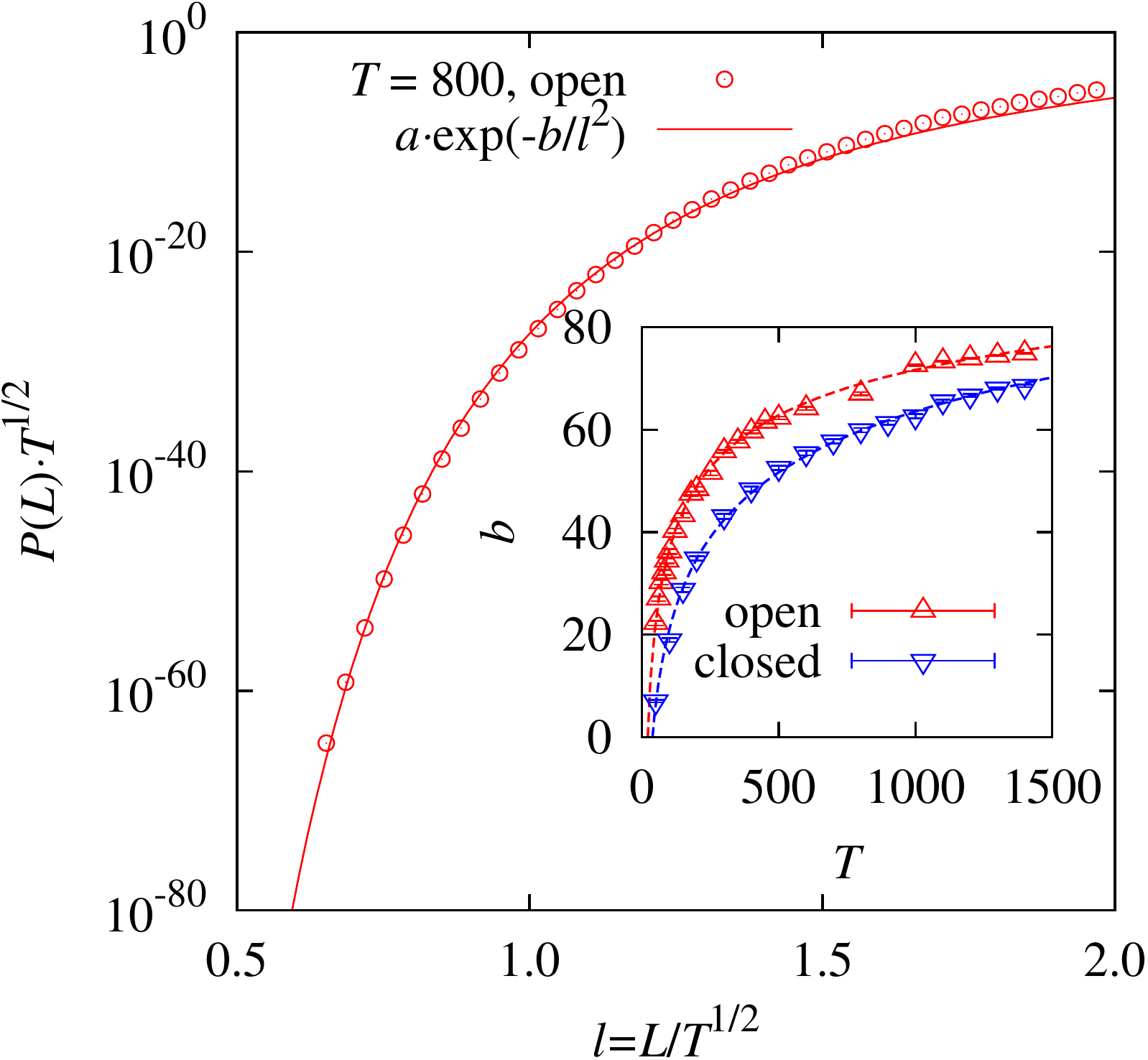} }
\caption{(color online) 
Main plot: rescaled distribution (symbols) of the perimeter for the 
open Gaussian walks for small values of the perimeter, length $T=800$
together with a fit to the function $ae^{-b/(L^2/T)}$.
Inset: Scaling behavior of the parameter $b$ as function of walk length,
for open and closed Gaussian walks, respectively. The lines represent
fits to functions of the form $b_{\infty}+c b^{\eta}$, resulting
in $b_{\infty}=129(9)$ (open) and $b_{\infty}=250(50)$ (closed),
with a very slow convergence ($\eta=-0.21(3)$ and $\eta=-0.09(2)$).
\label{fig:leftflank_peri}}
\end{figure}

Although the focus of our simulations was on the right tails
of the distributions, i.e., the large range of support for larger than
typical values, we have also performed some simulations to 
study very small values of the perimeter, to verify scaling
function (\ref{small_x}) leading to an expected $\sim e^{-b/(L^2/T)}$
behavior.

 The main plot of Fig.\ \ref{fig:leftflank_peri}
shows, as example, 
the resulting rescaled distribution for open Gaussian walks of length
$T=800$ together with a  fit to this functional form.  
Apparently, the data follows the predicted form very well in the
range of small perimeters.
 We have furthermore studied  the asymptotic behavior of the factor $b$
in the exponent and found (see inset of Fig.\ \ref{fig:leftflank_peri})
that the behavior is compatible with a convergence to a limiting value.

\section{Conclusions \label{sec:conclusions}}
We succeeded with the application of the above explained large-deviation 
scheme to the study of properties of convex hulls of random walk models.
We could obtain the corresponding distributions 
over broad ranges of convex hull area $A$ and perimeter $L$,
down to densities and probabilitiues as small as $10^{-300}$.

 The 
scaling behavior of these distributions turned out to be linked to 
walk length $T$ in the same way as the averages $\langle A \rangle \sim T$ and 
$\langle L \rangle \sim \sqrt{T}$. Also, the examination of the empirical 
rate functions resulted in simple power laws, which are compatible
with the scaling behavior. This leads to the natural assumtion
that the perimeter $L$ and the square root $\sqrt{A}$ of the area
are Gaussian distributed. Indeed either peak region of the data or
the tails region fit this form very well, but it is not possible
to fit over the full range of the support one single function.
Nevertheless, for almost all our results we found asymptotic agreement between
the Gaussian and the lattice walk case, as expected.

Since the application of large-deviation simulation approaches for this
problem turned out to be very useful, it should be applied
for similar problems. For future work, 
besides obvious extensions like considering higher
dimensions, we want to consider other random walk 
models, particularly multiple interacting 
walkers~\cite{Acedo2002,Kundu2014}, 
in simple self-avoiding walks as well as in loop-erased 
random walks\cite{Lawler1987,Majumdar1992} and, most 
importantly, for applications to biology like the formation of animal 
territories\cite{Giuggioli2011}.


\begin{acknowledgments}
We acknowledge personal support by O. Melchert and 
M. Manssen. Our simulations were carried out on the HERO cluster of 
Carl-von-Ossietzky Universit\"at Oldenburg, which has been funded by 
the DFG through its Major Research Instrumentation Programme 
(INST 184/108-1 FUGG) and the Ministry of Science and Culture (MWK) of 
the Lower Saxony State.
SNM acknowledges support
by ANR grant 2011-BS04-013-01 WALKMAT and in part by
the Indo-French Centre for the Promotion of Advanced Re-
search under Project 4604-3.

\end{acknowledgments}



{}

\end{document}